# New Directions for Thermoelectrics: A Roadmap from High-Throughput Materials Discovery to Advanced Device Manufacturing

*Kaidong Song, A. N. M. Tanvir, Md Omarsany Bappy, and Yanliang Zhang**

K. Song, A. N. M. Tanir, M. O. Bappy, Y. Zhang
Department of Aerospace and Mechanical Engineering
University of Notre Dame,
Notre Dame, IN 46556, USA
E-mail: yzhang45@nd.edu

## Abstract

Thermoelectric materials, which can convert waste heat into electricity or act as solid-state Peltier coolers, are emerging as key technologies to address global energy shortages and environmental sustainability. However, discovering materials with high thermoelectric conversion efficiency is a complex and slow process. The emerging field of high-throughput material discovery demonstrates its potential to accelerate the development of new thermoelectric materials combining high efficiency and low cost. The synergistic integration of high-throughput material processing and characterization techniques with machine learning algorithms can form an efficient closed-loop process to generate and analyze broad data sets to discover new thermoelectric materials with unprecedented performances. Meanwhile, the recent development of advanced manufacturing methods provides exciting opportunities to realize scalable, low-cost, and energy-efficient fabrication of thermoelectric devices. This review provides an overview of recent advances in discovering thermoelectric materials using high-throughput methods, including processing, characterization, and screening. Advanced manufacturing methods of thermoelectric devices are also introduced to realize the broad impacts of thermoelectric materials in power generation and solid-state cooling. In the end, this paper also discusses the future research prospects and directions.

## 1. Introduction

Since the discovery of Seebeck, Peltier, and Thomson effect in the 19$^{th}$ century, thermoelectric (TE) materials have attracted interest among scientists and engineers due to the profound merit of TE materials in building an energy-efficient world.[1–5] TE materials can generate electrical energy from a temperature gradient and vice versa. While two-thirds of the worldwide energy consumption is getting wasted as heat, TE devices can be a potential solution to improve energy efficiency by harvesting waste heat.[2] TE devices do not require moving parts or environmentally harmful working fluid which can provide a scalable and environmentally friendly power generation and cooling solution. The growing interest and research investment in this field have empowered a wide range of applications of TE devices in power generation in spaces and other remote locations, automotive and industrial waste heat recovery, as well as solid-state temperature controllers (such as car climate control, small portable cooler, microelectronics thermal management, etc.) aiming to replace the vapor compression based mechanical refrigerators.[6–11]

A TE device (TED) requires connecting the n-type and p-type semiconductor materials electrically in series and thermally in parallel as shown in **Figure 1**a. Depending on the application, TEDs can be of two major types – electricity generation (TEG) devices and cooling (TEC) devices. **Figure 1** demonstrates different kinds of TED architectures based on different applications and connection configurations. The figure of merit (ZT) is used to evaluate thermoelectric materials' efficiency, and the heat-to-electricity conversion efficiency (η) and coefficient of performance (COP) are used for assessing the performance of TEG and TEC respectively. The figure of merit is defined as $ZT = \frac{\alpha^2 \sigma T}{\kappa}$, where α is the Seebeck coefficient, $\sigma$ denotes the electrical conductivity, $\kappa$ is the thermal conductivity, and $T$ is the absolute temperature these parameters are measured at.[12] The $\alpha^2\sigma$ is defined as the power factor of the TE material,[8] which depends on the carrier concentration ($n$) and carrier mobility ($\mu$) of the material. These TE properties are temperature dependent which result in a peak ZT at certain temperature. Thereby a measure of average $ZT$ is used to evaluate the TE performance of a material over a wide temperature range. The heat-to-electricity conversion efficiency (η) is closely related to the average $ZT$ (denoted as $ZT_m$). The maximum conversion efficiency can be expressed as [13]

$$\eta_{max} = \frac{T_H - T_C}{T_H} \frac{\sqrt{1+ZT_m}-1}{\sqrt{1+ZT_m}+\frac{T_C}{T_H}} \quad (1)$$

where $T_H$ and $T_C$ are the hot side and the cold side temperatures of the device respectively. TE materials need a $ZT_m$ value close to 4 to reach an efficiency comparable to the current vapor compression based energy conversion technologies.[1,14] Most of the TE research in recent decades focuses on the improvement of material $ZT$ over a wide range of temperatures.[15–18] Although significant progress has been made in improving TE materials, the average $ZT_m$ and device performances still need further improvement.[19,20]

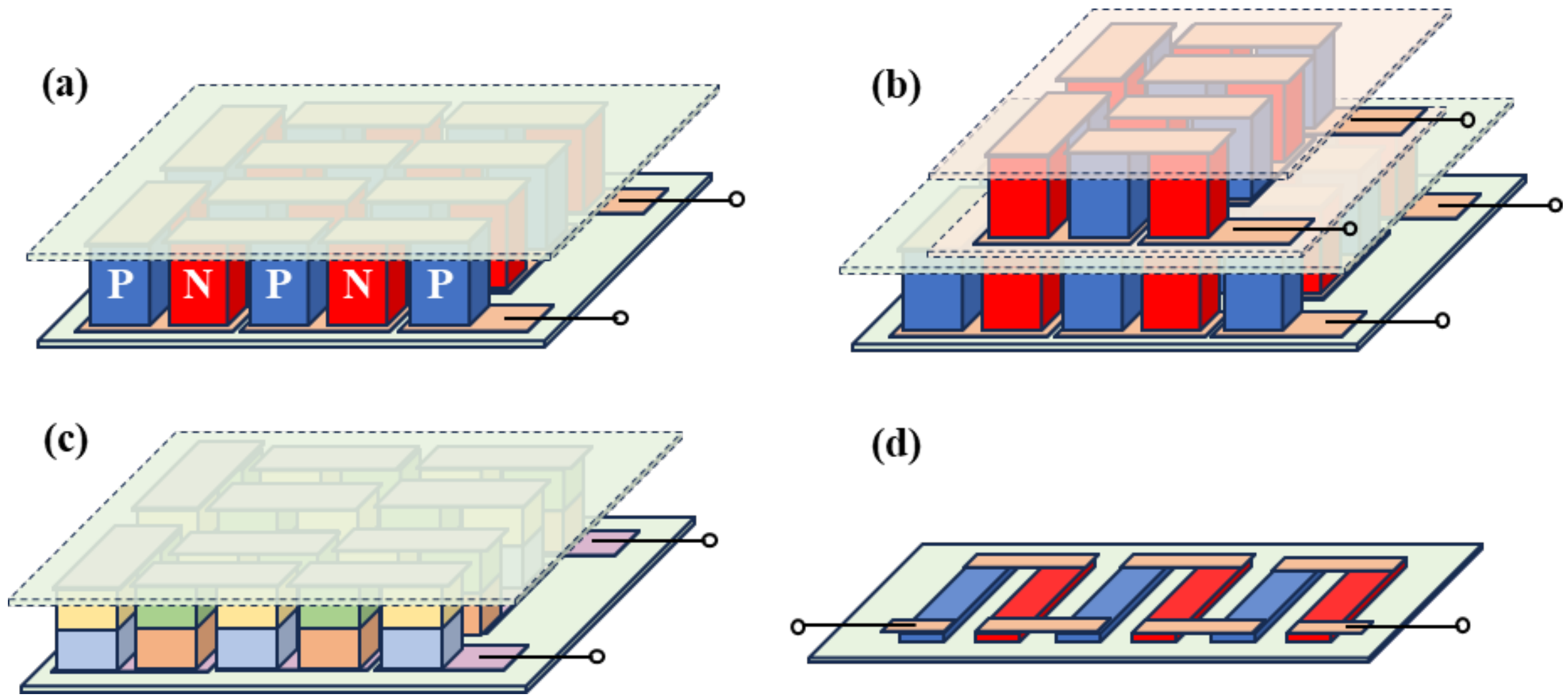

**Figure 1.** Different types of TED architecture (a) cross-plane devices with cold and hot sides in different planes of position corresponding to the TE legs, (b) cascaded device for utilizing large temperature difference and having large cooling output, (c) segmented device for efficiency improvement. Here, different color denotes different TE materials, (d) in-plane device configuration with the hot and cold side taking the same plane of the TE leg.

The field of high-throughput material discovery is emerging and has demonstrated its potential in discovering new energy materials.[21,22] By combining sophisticated material processing and characterization techniques with machine learning, scientists can create databases and evaluate large amounts of data to discover new materials with desired properties.[21] High-throughput materials synthesis and processing combined with machine learning provide unprecedented opportunities to expedite the discovery of new TE materials with enhanced performances.[22,23] Besides improving the materials' ZT, another important aspect of TE technology is to realize scalable and cost-effective manufacturing of TE materials and devices. Conventional TE material and device fabrication requires numerous processing steps, which are labor-intensive and costly. [20,24–27] Recent advancements in the scalable printing of TE inks provide a low-cost and highly scalable avenue to convert TE materials into microscale to macroscale TEDs directly.[28,29]

This paper will review recent progress in the high throughput discovery of TE materials and related development of high throughput material synthesis, processing, and characterization methods. We will also discuss high-throughput TED manufacturing using efficient, scalable, and low-cost printing and sintering methods, and various applications of TEDs. Lastly, the perspective on future research opportunities and directions will be briefly discussed for continuing progress toward efficient and cost-effective TEDs and realizing their broad applications.

## 2. High-throughput Materials Discovery

Benefiting from the high-throughput combinatorial materials discovery approach, the discovery of demanded thermoelectric materials has developed rapidly.[22,30] High-throughput combinatorial materials discovery begins with synthesizing or processing a diverse library of TE materials using techniques like magnetron sputtering, inkjet printing, and material extrusion.[22] Materials are prepared as thin films or bulk materials with gradient compositions, or an array of discrete samples with unique compositions. The TE properties of these materials such as thermal conductivity, electrical conductivity, and Seebeck coefficient are then rapidly characterized with high throughput.[31] The large amount of data generated during the characterization process will be further analyzed using big data analytics techniques and machine learning algorithms to reveal patterns and correlations to understand the relationship between the composition, structure, and TE properties of the material to screen out TE materials with the desired properties.[21,32] This high-throughput combinatorial approach significantly expedites the discovery and optimization of TE materials, fostering advances in TE technology and broadening its applications.[33]

### 2.1 High-throughput Combinatorial Materials Processing

The high-throughput combination method has been applied to the processing and manufacture of various materials.[30] Significant advances have also been made in the field of TE materials using these high-throughput material processing techniques. The following sections will delve into the various high-throughput combination processes used to process TE materials.

*2.1.1 High-throughput Combinatorial Printing from Aerosols*

In situ mixing and printing in the aerosol phase offer instantaneous adjustments to the mixing ratio of a wide array of materials on the fly, a feature not achievable with conventional multi-material printing utilizing liquid-liquid or solid-solid phases.[34,35] Zhang's group introduced a high-throughput combinatorial printing (HTCP) approach that can fabricate materials with microscale compositional gradients.[22] This method begins by converting two or more inks into aerosols comprised of micro-sized ink droplets. These streams of ink are then combined within a single nozzle and directed using a surrounding sheath gas prior to deposition (**Figure 2**a). This technique can deliver intricate designs with a horizontal resolution of approximately 20 μm and a deposition thickness close to 100 nm (**Figure 2**b). To showcase the versatility of the HTCP method, films with gradient compositions were produced using a wide variety of materials. These materials encompass chalcogenides, halides, nitrides, carbides, oxides, and metals, and include elements from both the periodic table's s- and p-block (**Figure 2**c). They further demonstrated this method in high-throughput discovery and optimization of TE materials. A $Bi_2Te_{2.7}Se_{0.3}$ film with gradient sulfur-doping was printed using HTCP to identify the optimal sulfur doping level that yields the highest TE power factor. As the sulfur doping concentration increases, the Seebeck coefficient of the printed $Bi_2Te_{2.7}Se_{0.3}$ film dramatically rose from −130 to −200 $\mu V K^{-1}$ (at roughly 0.5% S) before settling around −213 $\mu V K^{-1}$ (at approximately 1.0% S; **Figure 3**a). This notable change might be attributed to the rise in the density of states (DOS) effective mass brought about by sulfur doping.[36] The maximum electrical conductivity (refer to **Figure 3**b) and power factor (see **Figure 3**c) of the combinatorially doped $Bi_2Te_{2.7}Se_{0.3}$ film are achieved at an ideal sulfur concentration of roughly 0.6 and 1.0 atomic weight percentage (at.%), respectively. The focus of this printed material library is on discovering the best doping composition rather than target-specific properties. Considering the limitations in electrical conductivity for aerosol jet-printed films due to ink viscosity requirements, a more concentrated $Bi_2Te_{2.7}Se_{0.3}$ ink was used to create thicker films through extrusion printing for practical uses. Subsequent measurements of TE properties showed a peak room temperature power factor of 1,774 $\mu W/mK^2$ with 1.0% sulfur doping (as seen in **Figure 3**d). This value is notably higher than that of many other printed n-type TE materials, as illustrated in **Figure 3**e. These results underscore the HTCP's ability to accelerate the materials' screening and optimization process to attain desired properties.

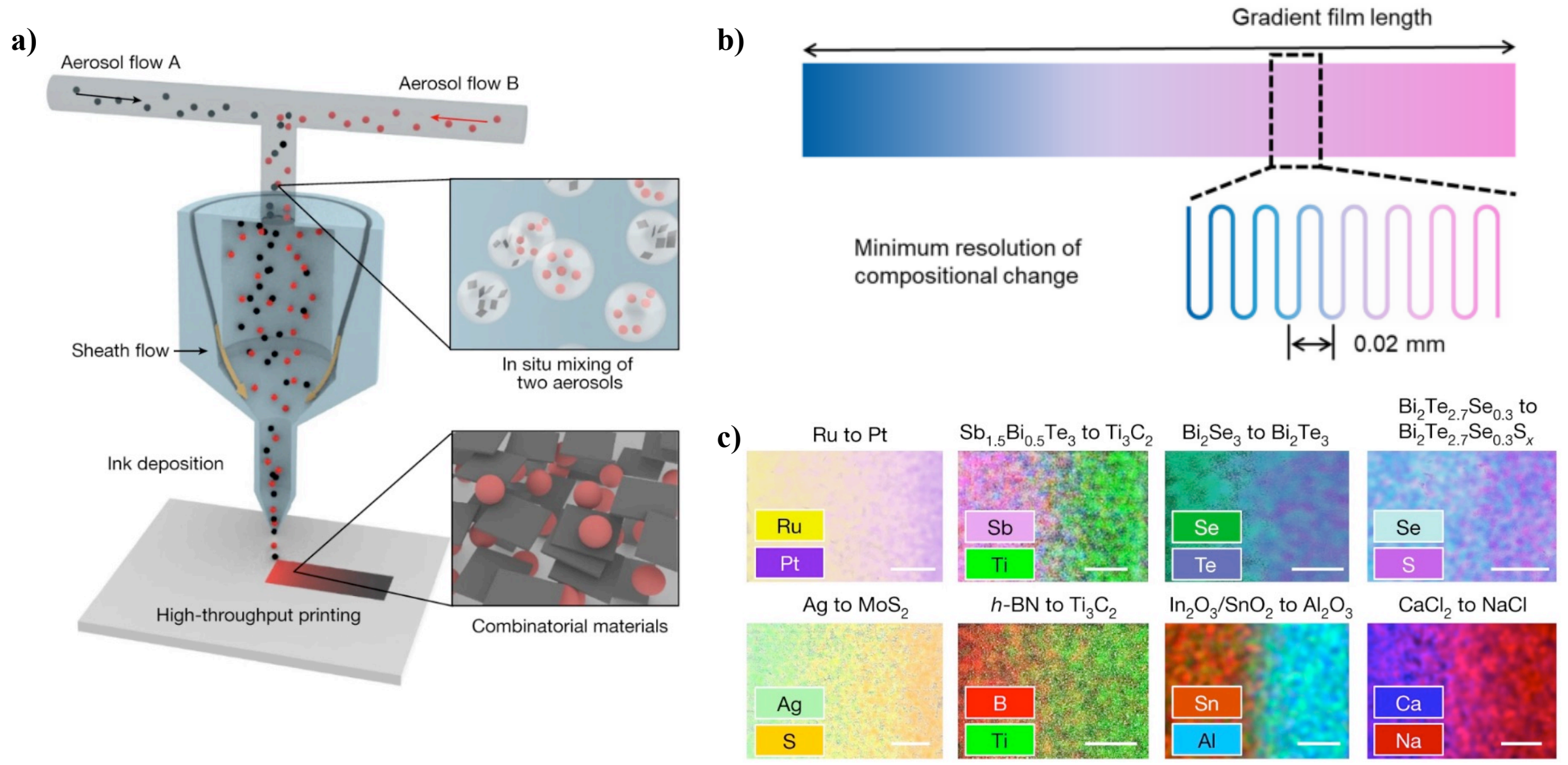

**Figure 2.** High-throughput combinatorial printing from aerosols. (a) Schematic representation of the combinatorial printing approach based on in situ aerosol mixing. (b) Schematic demonstration of the gradient design and printed pattern. (c) Elemental distribution visualization of diverse combinatorial prints spanning a wide array of elements. Scale bars, 300 μm. Reproduced with Permission.[22] Copyright 2023, The Authors. Springer Nature.

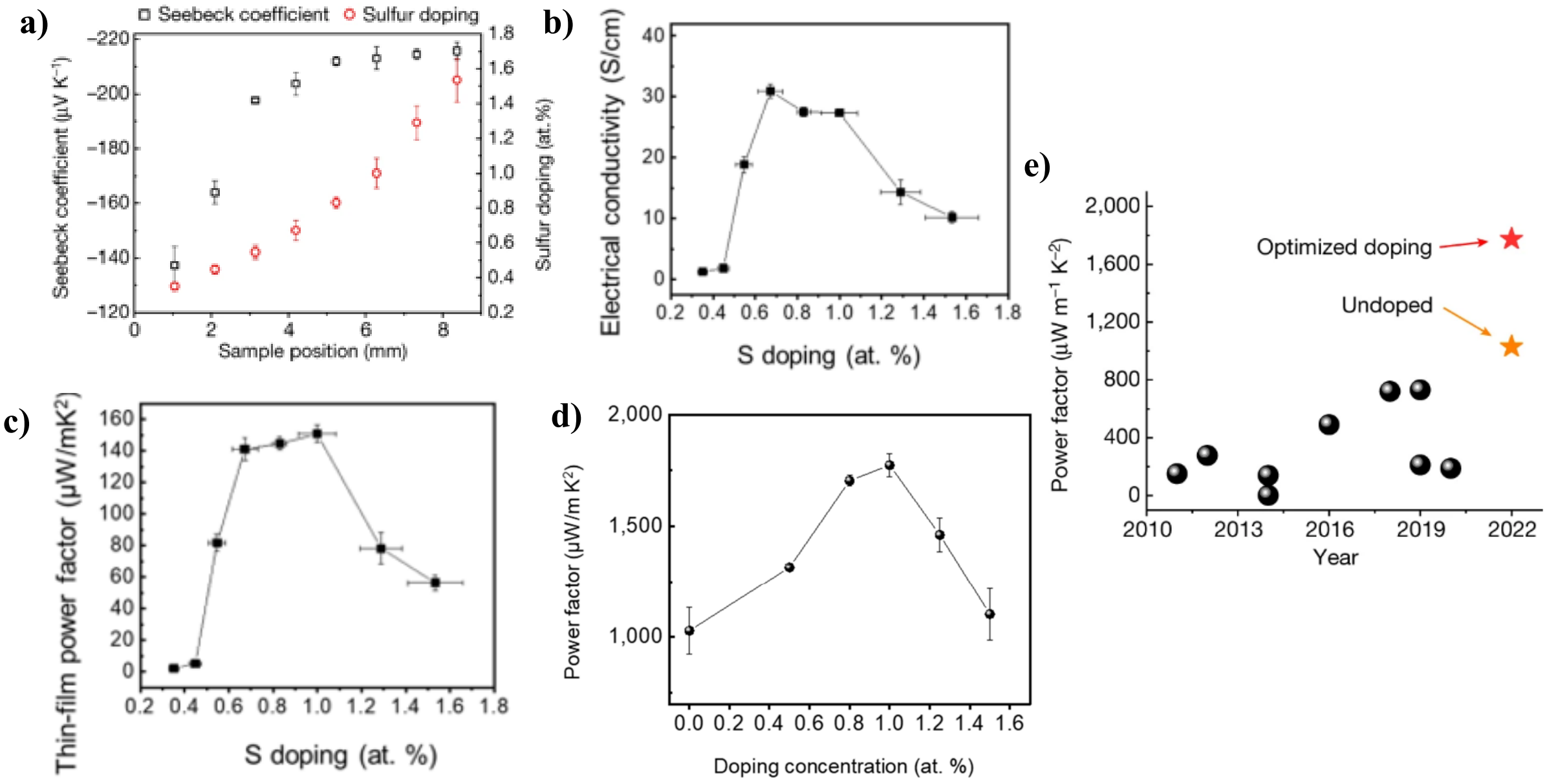

**Figure 3.** High-throughput discovery and optimization of $Bi_2Te_{2.7}Se_{0.3}$. (a) A $Bi_2Te_{2.7}Se_{0.3}$ film showcasing varying sulfur doping levels and the corresponding local variations in the Seebeck coefficient. (b) Electrical conductivity vs. sulfur doping concentrations. (c) Thin-film power factor vs. sulfur doping concentrations. (d) The TE power factor of extrusion printed $Bi_2Te_{2.7}Se_{0.3}$ samples with various sulfur doping concentrations. (e) Room temperature power

factor comparisonss over the past decade. Reproduced with Permission.[22] Copyright 2023, The Authors. Springer Nature.

### 2.1.2 High-throughput Material Co-Sputtering

High-throughput combinatorial sputtering method has been frequently used to accelerate the creation and identification of new materials across various compositions. In this process, the materials targeted are precisely adjusted by managing the deposition condition of the radio frequency (RF) power to achieve diverse material compositions.[37] This technique has been utilized by Goto et al. to create both p- and n-type bismuth telluride ($Bi_2Te_3$) thin films, and **Figure 4**a illustrates the equipment setup for the combinatorial sputter coating. During this process, the thin films' crystal structure and preferred orientation were altered by managing the RF power. At elevated RF powers, the rate of sputtering for Te exceeded that of Bi, allowing for precise control of the Bi/Te ratio in the films through RF power adjustments. After assessing TE properties, the p- and n-type films were optimized along with their ZT. The highest ZT values observed for the n- and p-type bismuth telluride thin films were 0.27 and 0.40, using RF powers of 90W and 120W, respectively.

Han et al. also introduced a high-throughput method for constructing a $Bi_2Te_{3-x}Se_x$ thin film library using this combinatorial sputter coating.[38] As illustrated in **Figure 4**b, the $Bi_2Te_3$ target and Te target were employed to create the Bi–Te film, while the $Bi_2Se_3$ target was utilized to add the Se element. This method essentially preserves a 3:2 ratio of anions to cations in the material. The three targets were placed 120° apart and directed towards the AlN substrate situated at one corner of the sample stage, allowing for a wide-ranging gradient of Se variation. A single preparation batch yielded a sample with 196 distinct compositions, laid out in a 14 × 14 matrix. This high-throughput technique allows for the simultaneous exploration of TE thin film compositions and microstructures with just 2–3 preparation batches. High-performance $Bi_2Se_{0.1}Te_{2.9}$ films were achieved, with an impressive average ZT value of 1.047 in the range of 313 to 523 K and a peak ZT value of 1.303 at 353 K.

### 2.1.3 High-throughput Combinatorial Pulsed Laser Deposition

Recent progress in the field of combinational material deposition technologies has accelerated the development of new materials. Among the many methods, combinational pulsed laser deposition (PLD) has been identified as a dependable tool for synthesizing TE materials. [39–41]

This approach enables the rapid creation of thin film samples of diverse compositions. What sets PLD apart from other thin-film deposition techniques like chemical vapor deposition or molecular beam epitaxy is its ability to accurately replicate the intricate stoichiometries from the target material onto the deposited layer.[42] Additionally, PLD introduces unique nanostructuring features within thin films, such as dimensional reduction and texturing, which help improve TE performance.[43] As depicted in **Figure 4**c, combinatorial PLD entails sequentially depositing small quantities (less than a unit cell) of each target material. Typically, after a complete rotation across all the targets, the material deposited over the combinatorial spread area achieves a thickness close to one unit cell. This procedure is repeated until the target thickness is attained.

In one study, Snyder and colleagues leveraged this high-throughput combinatorial PLD technique to optimize TE thin films made of complex materials like layered misfit cobaltite, $Ca_3Fe_xCo_{4-x}O_9$. [41] In another instance, Watanabe and the team employed the PLD process to synthesize and characterize TE intermetallic compounds of $Mg_xSi_yGe_{1-y}$, adjusting the compositions (x = 2.3–4.1, y = 0.45–1).[40] By analyzing the specific composition of the deposited thin film samples, it is confirmed that the PLD process enabled the successful creation of proper compositional gradients within the library. The entire process, including the deposition of amorphous precursors, annealing, examination by electron probe microanalysis and XRD, and evaluation of Seebeck coefficient and electrical resistivity, took 12.5 hours, demonstrating the method's high throughput nature.

*2.1.4 High-throughput Wire-fed Droplet Alloying*

Different from the above-mentioned film-based combinatorial processing methods to expedite the identification of novel and efficient TE materials, Garcia and colleagues pioneered a combinatorial method to generate a vast array of bulk alloy samples using a wire-fed suspended droplet alloying (SDA) process and different pure element wires (Refer to **Figure 4**d).[44] This method is based on a laser beam melting the raw material of the wire, creating the molten core of the bulk alloy. By modulating the feed rates of the wires, it's possible to forge samples of differing compositions. During the formation of the alloy, the laser melts the aligned wires, creating droplets that grow and combine into one alloy droplet, held up by surface tension. When it reaches a certain mass, it descends onto a substrate. Additional droplets are layered, with the lower part solidifying, and a molten upper section. The sequence is repeated until the necessary thickness is achieved, followed by cooling under a shield of argon gas.

The efficiency of this accelerated technique was illustrated through the successful identification of an efficient TE ternary system, Al–Fe–Ti. In this approach, approximately 500 samples, including single elements and binary and ternary alloys, were first created, and their power factor was examined. A notable measurement included the maximum Seebeck coefficient value of −57 µV/K, linked to the $Al_{12.5}Fe_{37.5}Ti_{50}$ composition, and electrical resistivity values in the magnitude of $10^{-4}$ Ω cm. The peak power factor of $13.3 \times 10^{-4}$ W/m $K^2$ was found in the same composition, while the highest power factor ($7.0 \times 10^{-4}$ W/m $K^2$) for samples with a positive Seebeck coefficient was seen in $Al_{30}Fe_{55}Ti_{15}$. The discovery of a high-power factor in the Al–Fe–Ti system affirms the applicability and effectiveness of this high-throughput method in advancing TE material discovery.

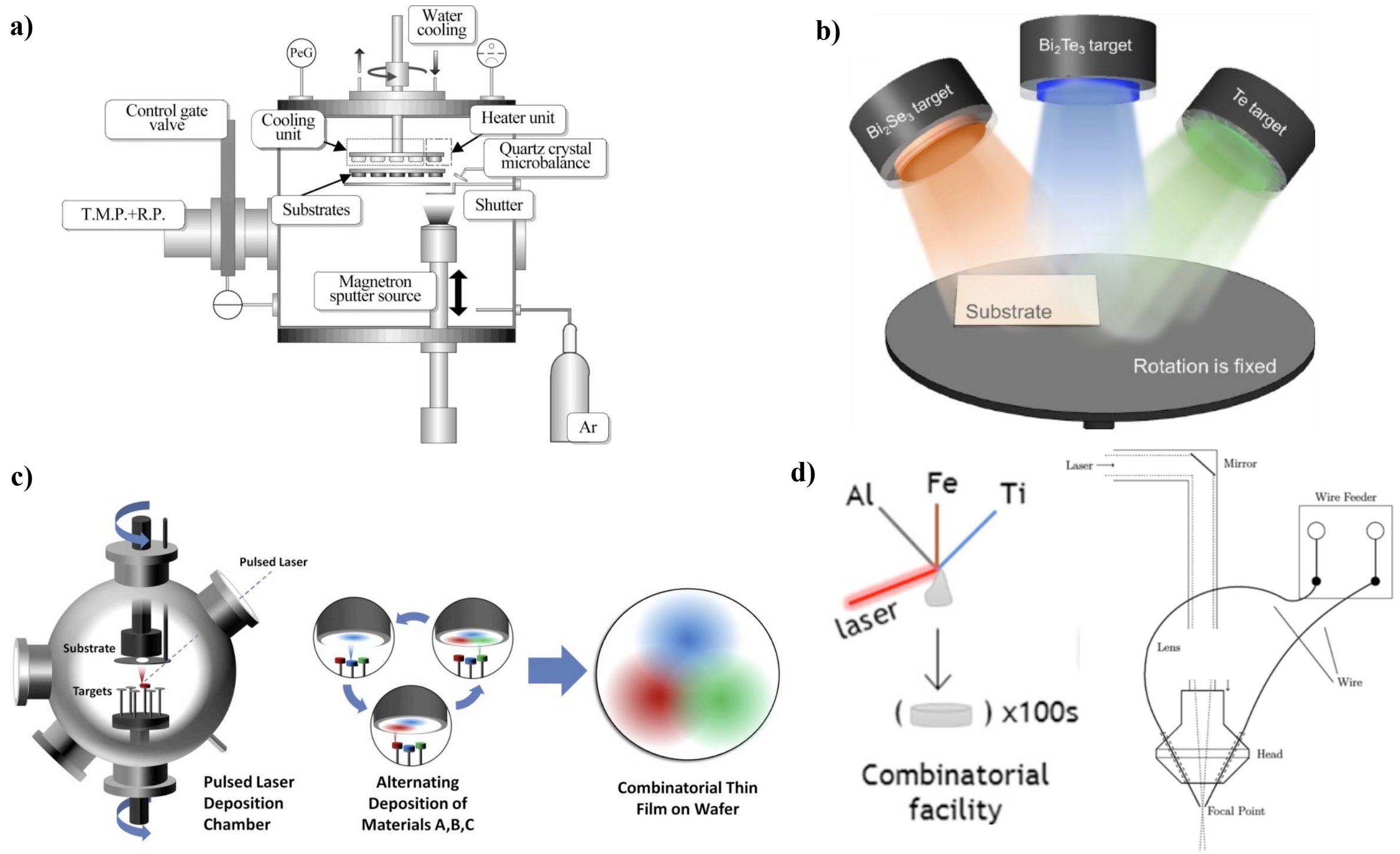

**Figure 4.** High-throughput combinatorial processes. (a) Schematic representation of the experimental arrangement for the combinatorial sputter coating system. Reproduced with Permission.[37] Copyright 2017, Elsevier. (b) Thin films of Bi-Te-Se with diverse compositional gradients produced using combinatorial sputter coating. Reproduced with Permission.[38] Copyright 2023, Elsevier. (c) Schematic of combinatorial pulsed laser deposition. Reproduced with Permission.[41] Copyright 2015, Elsevier. (d) Schematic illustrating the wire-fed suspended droplet alloying method. Reproduced with Permission.[44] Copyright 2016, ACS Publications.

### 2.2 High-throughput Material Characterization and Screening

Combinatorial materials science crucially relies on combinatorial materials characterization and screening. This process involves examining numerous material compositions and properties in a systematic and high-throughput manner, significantly accelerating the discovery and optimization of materials. In particular, computational characterization and screening, which employs advanced computational models and algorithms to evaluate and predict the TE properties of a vast array of materials, have played an essential role in the hunt for new TE materials.[21,45]

#### 2.2.1 High-throughput Material Characterization

In high-throughput material characterization approaches, it is crucial to develop methods that can efficiently evaluate the performance of TE materials within a combinatorial library.
A TE screening instrument was developed by Tang's group to allow temperature-dependent mapping of the Seebeck coefficient and electrical resistivity within 300 K to 800 K.[46] **Figure 5**a showcases the tool, highlighting the probe used to characterize the Seebeck coefficient. This hot probe consists of an R-type thermocouple housed within a small heater, complemented by a second thermocouple acting as a temperature sensor on the cold side. Sheet resistance is determined using four tungsten wires placed in a ceramic tube, all connected to a spring mechanism for individual motion. The measurement process takes roughly 30 seconds per point, or approximately 2.5 hours, for the whole film with a diameter of 76.2 mm at each furnace temperature. This screening instrument effectively produces a Seebeck coefficient profile for a ternary $CoSb_3$-$LaFe_4Sb_{12}$-$CeFe_4Sb_{12}$ film (as shown in **Figure 5**b) and a sheet resistivity profile (illustrated in **Figure 5**c). With optimized conditions, the proposed screening tool yields a sheet resistivity and Seebeck coefficient measurement uncertainty of less than 10% and 5%, respectively. Another measurement system for scanning thermal conductivity was devised using the frequency domain thermoreflectance method in conjunction with periodic heating.[31] This technique is based on the correlation between changes in a material's optical reflection coefficient and temperature changes. The setup comprises a compensating network, an XYZ-axis sample stage controlled by a motor, a voltage-driven heater, and two laser diodes (**Figure 5**d). The specimen, a TE film with a thin molybdenum layer, is selectively heated using a modulated laser. A second laser detects the film's thermal response. By calculating the surface temperature response between the heated laser indicators and the thermoreflectance, the thermal effusivity (b) can be obtained, which is related to the thermal conductivity as $b = (\kappa c_p \rho)^{1/2}$. Once b is obtained experimentally, the thermal conductivity (κ) can be easily derived. **Figure 5**e displays the two-dimensional data set of thermal conductivity values for a Si sample.

Zhang et al. presented a non-contact scanning thermal microprobe that assesses thermal conductivity (κ) with a spatial resolution of 3 μm. This method utilizes quasi-ballistic air conduction across a 10-100 nm air gap between a microprobe heated by the Joule effect and the sample.[47] The microprobe features a Veeco® Wollaston wire with a 5-μm Pt–Rh core, enveloped by a 75-μm silver shell, acting both as a heater and a thermistor (**Figure 5**f). The temperature rise $\Delta T_{probe}$ of the wire is calculated from the calibrated electrical resistance and temperature coefficient of resistance and increases linearly with the input power. Compared with traditional methods, the κ value deviation obtained by this device is only 5% to 10%. This contactless microprobe reduces the influence of heat transfer on the surface chemistry and morphology of the sample, making thermal characterization of various nanostructured materials possible.

Zhang's group has also developed a high-sensitivity scanning thermal probe capable of simultaneously measuring the Seebeck coefficient and thermal conductivity, extending the measurement range of thermal conductivity up to 18 W/m K. This range improvement allows for the application of scanning thermal microscopy (SThM) to a broader array of samples compared to traditional commercial probes with reduced sensitivity above approximately 10 W/m K.[48] A distinctive feature of their design includes a microscale support structure made from a stiff, low-conductivity material that upholds the probe tip, ensuring that only the very end of the probe is exposed (as depicted in **Figure 5**g). This setup allows for the application of increased force between the probe and the specimen without harming the probe tip, more than doubling the magnitude, which reduces the thermal contact resistance between the probe and the sample by nearly 96%. This boosts the measurement sensitivity by upwards of 240% compared to comparable commercial probes working on the same concept. While taking measurements, an alternating current warms a section of the sample through the probe. The heat exchange between the probe and sample accounts for both the thermal contact resistance and the inherent resistance of the sample. By utilizing the probe's resistance and a pre-calibrated heat transfer model, the sample's thermal conductivity is ascertained. The Seebeck coefficient of the sample is then derived from the temperature rise and the recorded DC Seebeck voltage (refer to **Figure 5**h). This innovative probe is game-changing for its ability to measure thermal conductivity and Seebeck coefficients at the microscale level for materials that were previously not compatible with SThM. Furthermore, it facilitates a more rapid evaluation of combinatorial materials, streamlining the search for materials with specific TE traits.

Sasaki et al. have developed an integrated platform for evaluating the Seebeck coefficient, thermal conductivity, and internal strain of TE materials simultaneously, specifically for materials like bismuth telluride, whose TE properties can be modulated by inducing internal strain.[49] This approach combines scanning thermal probe microimaging (STPM) and micro point X-ray diffraction (XRD) (**Figure 5**i). During the test, bismuth telluride thin film (BTTF) samples are attached with high-temperature and water-cooled blocks at both ends. Radiation thermometers and thermocouple thermometers are then used to control and monitor the temperature of the sample. The STPM system and micro-XRD system were then used to evaluate the out-of-plane properties of the gradient-annealed BTTF samples, including crystal structure, Seebeck coefficient, and thermal conductivity. The internal stress of BTTF was characterized by micro-point XRD. The XRD patterns of the internal stress at different positions on the BTTF sample obtained are shown in **Figure 5**j.

To accelerate the measurement efficiency, Funahashi et al. developed a method to measure the Seebeck coefficients at 650 samples/h.[50] The measurement was conducted at ambient temperature using a two-terminal approach (as shown in **Figure 5**k). The Seebeck tester consists of two K-type thermocouples, with one connected to a heater to create a temperature differential across the sample ends. Depending on the DC supplied to the heater, the hot side's temperature can be adjusted between 300–340 K. The two terminals make contact with the opposite ends of the linear samples. Both the temperature variance and the TE voltage between the ends are recorded and sent to a computer through digital voltmeters. Otani et al. have also developed a high-throughput screening instrument that swiftly maps the TE power factors of libraries containing combinatorial composition-spread films.[51] This apparatus includes a probe that measures electrical conductivity and the Seebeck coefficient, coupled with an automated stage for probe navigation in three dimensions (X-Y-Z). The probe assembly consists of four spring-loaded probes, two thermometers, a heating element, a pair of insulators, and two copper plates (as shown in **Figure 5**l). To ensure accurate measurement of Seebeck's coefficient and reduce thermal resistance, one spring probe is fixed to each copper plate, while the other is fixed to the insulator. The probes are strategically arranged in a square with a gap of 1 mm between them. To maintain a consistent temperature, both the probe and the sample stage are housed in a protective housing. The standard four-probe method is used to measure the conductivity, while the Seebeck coefficient is determined by comparing the voltage and temperature difference between two points. These two parameters can measure each sample point in 20 seconds, allowing more than 1,000 measurements in less than 6 hours.

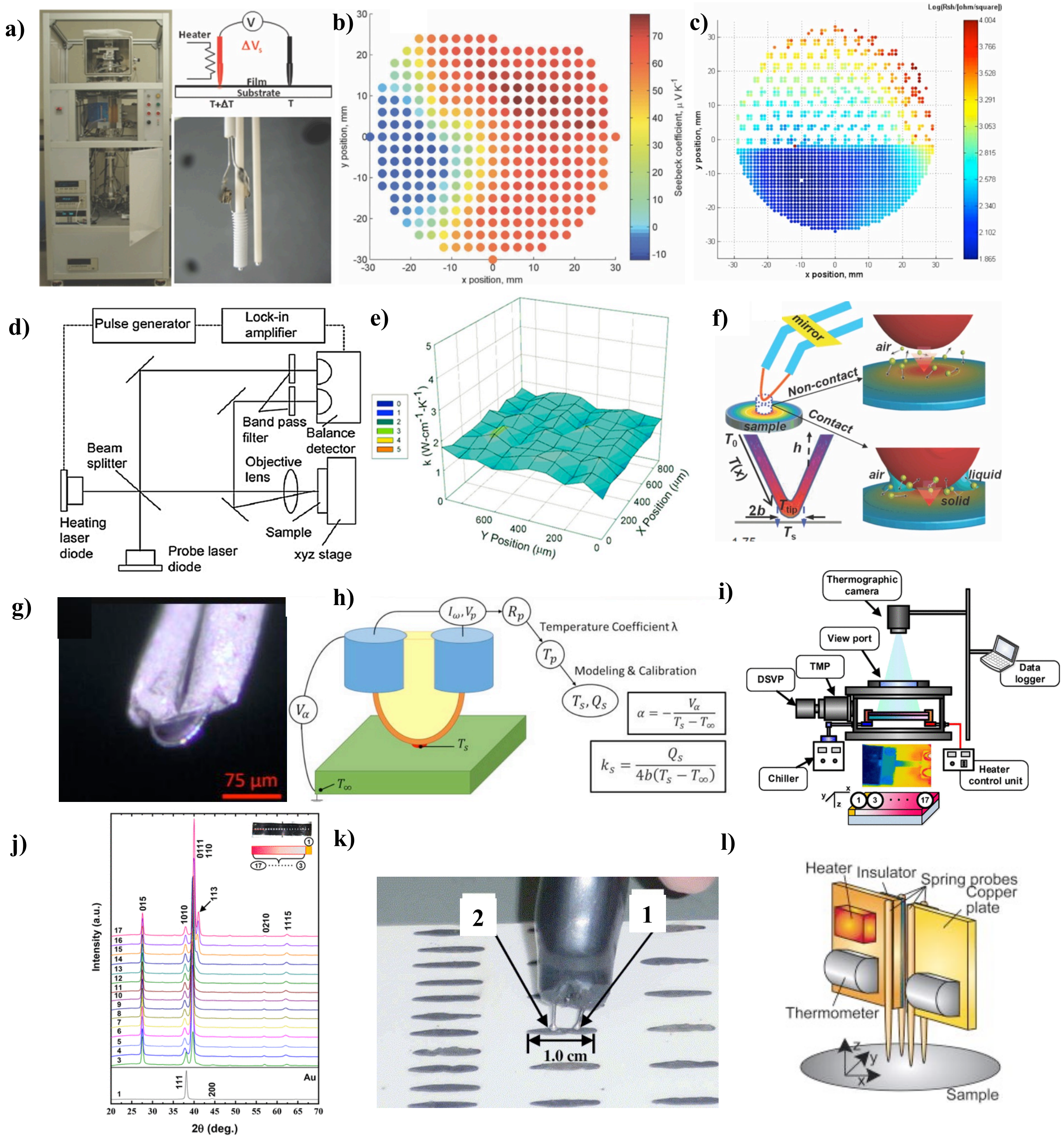

**Figure 5.** (a) An image of the high-temperature TE power factor screening instrument, accompanied by a depiction of the Seebeck coefficient probe (diagram at the top-right and actual photo at the bottom-right). (b) Contour plots of the Seebeck coefficient for a tri-component $CoSb_3$-$LaFe_4Sb_{12}$-$CeFe_4Sb_{12}$ combinatorial film. (c) Sheet resistance contour plot for the partially masked Sb film. Reproduced with Permission.[46] Copyright 2013, AIP Publications. (d) A schematic illustrating the scanning thermal effusivity frequency setup. (e) Data on the thermal conductivity of the Si sample. Reproduced with Permission.[31] Copyright 2015, Springer. (f) Schematic representations of the thermal microprobe, showcasing the routes of heat transfer between the microprobe tip and the sample in both contact and non-contact configurations. Reproduced with Permission.[47] Copyright 2011, The Authors. AIP Publications. (g) Microscope images of the custom-made thermal microprobe. (h) Schematic depicting the probe measurement principle. Reproduced with Permission.[48] Copyright 2021, The Authors. AIP Publications. (i) Schematic representation of the Combinatorial Gradient Thermal Annealing (COGTAN) method. (j) Micropoint XRD patterns of BTTF. Reproduced with Permission.[49] Copyright 2020, ACS Publications. (k) Assessment of the Seebeck coefficient at ambient temperature using the 'Seebeck tester'. Part 1 and Part 2 represent the hot and cold terminals, respectively, which are made up of K-type thermocouples. Reproduced with Permission.[50] Copyright 2004, Elsevier. (l) Schematic representation of the measurement probe designed for assessing both electrical conductivity and the Seebeck coefficient. Reproduced with Permission.[51] Copyright 2007, AIP Publications.

### 2.2.2 High-throughput Computational Screening

High-throughput computational screening is an effective and promising approach for discovering new TE materials with high performance. This strategy typically combines sophisticated thermodynamic techniques with electronic structural approaches, complemented by intelligent data analysis and database development, to take full advantage of contemporary supercomputing power to screen out desired materials.[21] Cases of fruitful high-throughput computational screening include examining chalcogenides. Layered IV-V-VI semiconductors have high prospects in TE applications due to their inherent ultra-low lattice thermal conductivity. Yet, gauging their TE efficacy using traditional trial-and-error experimental methods presents significant challenges. Gan et al. introduce a machine learn-based approach to accelerate the identification of potential TE materials in chalcogenides.[52] Using data sets obtained from high-throughput ab initio calculations, they constructed two neural network models that are good at predicting ZT peaks ($ZT_{max}$) and associated doping types. The standout material, n-type $Pb_2Sb_2S_5$, was pinpointed, achieving a $ZT_{max}$ greater than 1.0 at 650 K, which is attributed to its exceptionally low thermal conductivity combined with a respectable power factor. Jia et al. also utilized the high-throughput computational screening method to screen chalcogenides.[53] Using the deformation potential method within the single-band model, they determined the carrier relaxation time. They also formulated an electrical descriptor ($\chi$) that utilizes the effective mass of the carrier, allowing estimates of the peak power factor without the need to solve the electron Boltzmann transport equation. To efficiently gauge lattice anharmonicity and thermal conductivity, the Grüneisen parameter ($\gamma$) was calculated utilizing elastic properties, thus bypassing the lengthy phonon calculation. When the two descriptors were applied to binary chalcogenides, they evaluated 243 semiconductors and identified 50 with significant TE potential. These include TE compounds that have been identified, as well as 9 P-type and 14 N-type disulfide compounds that have not previously been associated with TE.

Recently, by adjusting the Cu dopant in hot-pressed AgSe-based TE composites, an electrical conductivity of approximately 9000 S/cm and a power factor of about 1600 $\mu W/mK^2$ were achieved at room temperature. Work is still ongoing to improve the TE characteristics of these materials. Shang and colleagues introduced a combined data-driven method that combines Gaussian Process Regression (GPR) and Bayesian Optimization (BO) to fine-tune the composition of five elements (Ag, Se, Te, S, and Cu) in AgSe-based TE materials.[23] As shown in **Figure 6**a, this approach consists of three cyclical steps: 1) gathering data from studies and experiments, 2) training with GPR and making predictions with BO, and 3) validating through experiments. As can be seen from the parity plot in **Figure 6b**, the power factor prediction of the initial GPR model based on literature data is in good agreement with the actual results. The model's effectiveness was affirmed by the close match between the predictions and actual power factors, as shown in **Figure 6**c. **Figure 6**d highlights the suggested compositions during each iteration. After seven iterations, the optimized AgSe materials exhibited a zT of 0.9 and a high power factor of 2100 $\mu W/mK^2$ at room temperature, marking a 75% enhancement from the initial composition (**Figure 6**e). Their study paves the way for utilizing active machine learning in expediting material system developments with fewer experiments.

In another investigation, a high-throughput framework was employed to study the electronic structures and p-type TE characteristics of diamond-like $ABX_2$ compounds. It not only helped identify compounds but also helped recognize underlying trends. [54] From nearly 85,000 entries in the Materials Informatics Platform, 65 were selected for examination. The analysis shows that the existence of a general conductive network in the anion X sublattice affects the electrical transport properties of the compound. Subsequent studies of 41 pnictide and chalcogenide compounds showed that pnictides exhibit lower Seebeck coefficients due to their smaller effective mass near valence band peaks. However, due to their increased electron group velocity and extended relaxation time, they have superior electrical conductivity and power factor. Utilizing the Slack model, 12 novel high ZT materials, both p-type and n-type in the $ABX_2$ format, were projected.

The TE properties of BTTF can also be modulated through a mixture of combined gradient thermal annealing (COGTAN) and machine learning, as the internal strain gradually changes with different annealing temperatures.[49] A design framework was established to determine the ideal internal strain of $Bi_2Te_3$ for superior TE performance, combining machine learning with experimental measurements, as shown in **Figure 7**a. The crystal structure (**Figure 5**j) and TE performance data (**Figure 7**b) collected through the combined characterization show the relationship between internal strain and TE properties. The Seebeck coefficient of BTTF fluctuates between 7.9 and -108 μV/K. The collected data was analyzed using artificial neural networks for machine learning to investigate in more depth the potential to improve Seebeck's coefficient. The established model shows that the optimum strain of 3-4% on the A-axis and 1-2% on the C-axis can significantly increase Seebeck coefficient.

High-throughput computational analysis has been employed to identify room-temperature Peltier cooling substances in Heusler compounds, as well as potential metal-organic frameworks for thermal conductivity.[55,56] Specifically, Luo et al. employed this technique to identify five promising Peltier cooling materials at room temperature: $GaSbLi_2$, $HgPbCa_2$, $SnTiRu_2$, $GeYbLi_2$, and $GeTiFe_2$, from a pool of 2958 Heusler compounds housed in the MatHub-3d database.[55,57] On a different note, aiming to discern the structural influences on heat transport in MOFs, Islamov and his team conducted extensive computational evaluations of thermal conductivity (κ) in MOFs. Utilizing classical molecular dynamics simulations on 10,194 theoretical MOFs generated by the ToBaCCo 3.0 software, they determined that MOFs with higher densities (exceeding 1.0 g $cm^{-3}$), smaller pores (less than 10 Å), and tetra-connected metal nodes are more likely to exhibit high thermal conductivity.[56] Interestingly, they identified 36 MOFs with ultra-low thermal conductivity (less than 0.02 W/mK), attributed mainly to their exceptionally large pore sizes (around 65 Å). Moreover, six of the investigated MOFs exhibited remarkably high thermal conductivity, exceeding 10 W/mK. Overall, the high-throughput computational screening method offers a faster and more systematic way to discover and optimize new TE and other functional materials with specific desired properties. Nevertheless, there are still opportunities and challenges ahead, especially concerning efficiency, accuracy, and the combination of machine learning techniques with experimental approaches to enhance material properties.[21,58]

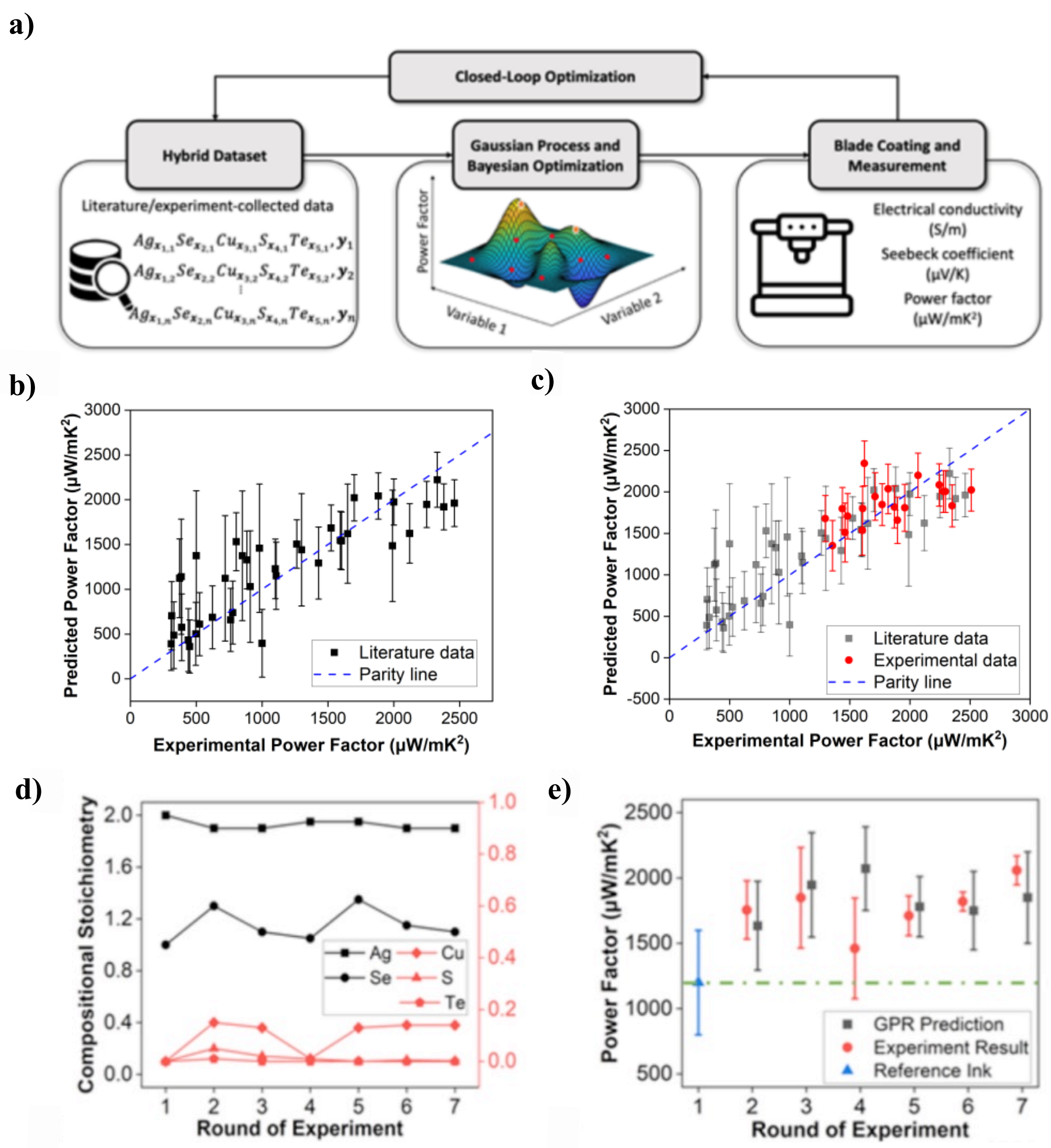

**Figure 6.** (a) Schematic representation of the closed-loop optimization through the cyclical process of machine learning and empirical data analysis. (b) A parity plot that displays power factors cited in scholarly sources (considered as the ground truth) juxtaposed with predictions from the initial GPR model, which was exclusively trained on the literature data. c) A parity plot that contrasts the experimental power factors with the predictions derived from the final GPR model, which was trained using the hybrid dataset. d) Chemical makeup of the materials produced during each experimental cycle. e) Power factors as determined through experimental measurements versus those predicted by the GPR model. Reproduced with Permission.[23] Copyright 2023, The Authors. Published by Wiley.

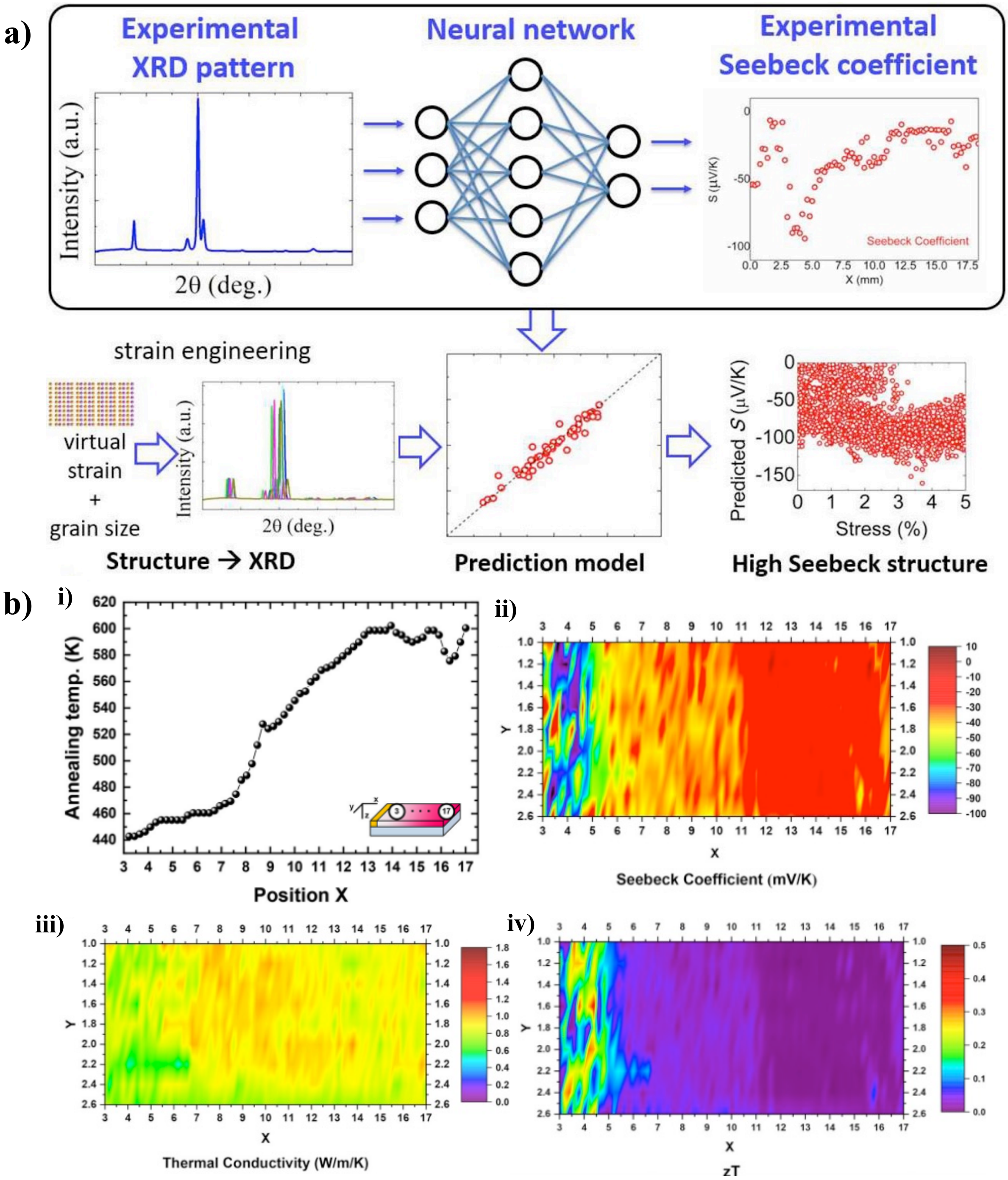

**Figure 7.** (a) Illustration depicting the application of XRD data in crafting TE materials through machine learning. (b) TE characteristics of the BTTF COGTAN sample derived from STPM mapping data: (i) Annealing temperature at specific sample locations, (ii) Seebeck coefficient, (iii) thermal conductivity, and (iv) projected zT value. Reproduced with Permission.[49] Copyright 2020, ACS Publications.

## 3. High-throughput Manufacturing of High-performance and Low-cost Thermoelectric Devices

The current cost of TEDs is still higher than other low-cost renewable energy technologies.[2] A large portion of the cost comes from the manufacturing of TEDs.[59] The innovation and scalability of ink-based printing can significantly reduce the cost of TEDs by reducing the materials and manufacturing costs. Ink-based printing combined with rapid post-printing processing and sintering can efficiently convert TE particles or precursors to devices with well-densified structures. Herein, we will first discuss different printing methods and their applicability in printed TE, followed by rapid sintering techniques and electrical contact processing techniques.

### 3.1 Ink-based Printing Techniques of Thermoelectric Devices

Several printing methods have been developed (e.g., screen printing, inkjet printing, aerosol jet printing, and extrusion printing) for printing functional devices.[60,61] In general, a colloidal suspension of the functional material is prepared first. The ink may contain several additives to make the ink printable, including a solvent, viscosity modifier, binder, defoamer, etc. A complete review of preparing inks for TEs has been provided by Zeng in 2022.[28] The prepared ink is deposited in a predetermined architecture enabling near-net shape manufacturing. Based on the procedures used ink-based printing can be applied to realize 2D thin film devices to 3D bulk device architectures. The near net shaping capability also enables complex and overhanging structures which is difficult to fabricate through conventional manufacturing processes. Moreover, precise control over ink composition for each layer makes it possible to print functionally graded structures. **Table 1** provides a summary of the ink-based printing methods and their properties.

**Table 1.** Common examples of ink-based printing techniques and their main attributes

| Printing Method | Ink viscosity (mPa s) | Printing speed (mm/s) | Printing Resolution | Advantages | Limitations |
|---|---|---|---|---|---|
| Screen Printing | 1000-10000 | N/A | ~30 μm[62][ | Cost-effective, high throughput | Limited resolution |
| Ink-jet Printing | 1-50 | ~5000[63] | ~10 μm[64] | High resolution, non-contact printing, protects the substrate against contamination or harm | Limited material selection for low viscosity inks, larger particle size can block nozzle |
| Aerosol Jet Printing | 1-1000 | 1~15[65] | ~10 µm[66] | High-resolution, non-contact printing, protects the substrate against contamination or harm, conformal printing | Limited material selection, complex ink formulation, low scalability |
| Extrusion Printing | 30-$10^7$ | ~500[67] | ~50 µm[68] | Easy to use, wide viscosity range | Limited resolution |

The following subsections will include descriptions of different printing processes and their application in printing thermoelectric devices focusing on inorganic materials, as there are already several excellent reviews on organic materials for TEs.[69–74]

### *3.1.1 Screen Printing*

Screen printing is one of the widely used processes for printing functional materials and devices such as electronics.[75–77] In this process, a preformed patterned grid or template of defined mesh (called squeegee **Figure 8**a) is used as a mask to print the specific shape and size by squeezing the ink out of the grid to the substrate. A blade can be used to perform the squeezing of the ink. The squeezed-out ink can be subsequently dried and sintered to achieve the desired density and property of the printed film. The printing thickness usually depends on the number of printed layers as well as the concentration of the ink.[78][79] The screen-printing process is scalable to different shapes and sizes without any moving parts or complex control systems, enabling low-cost applications.

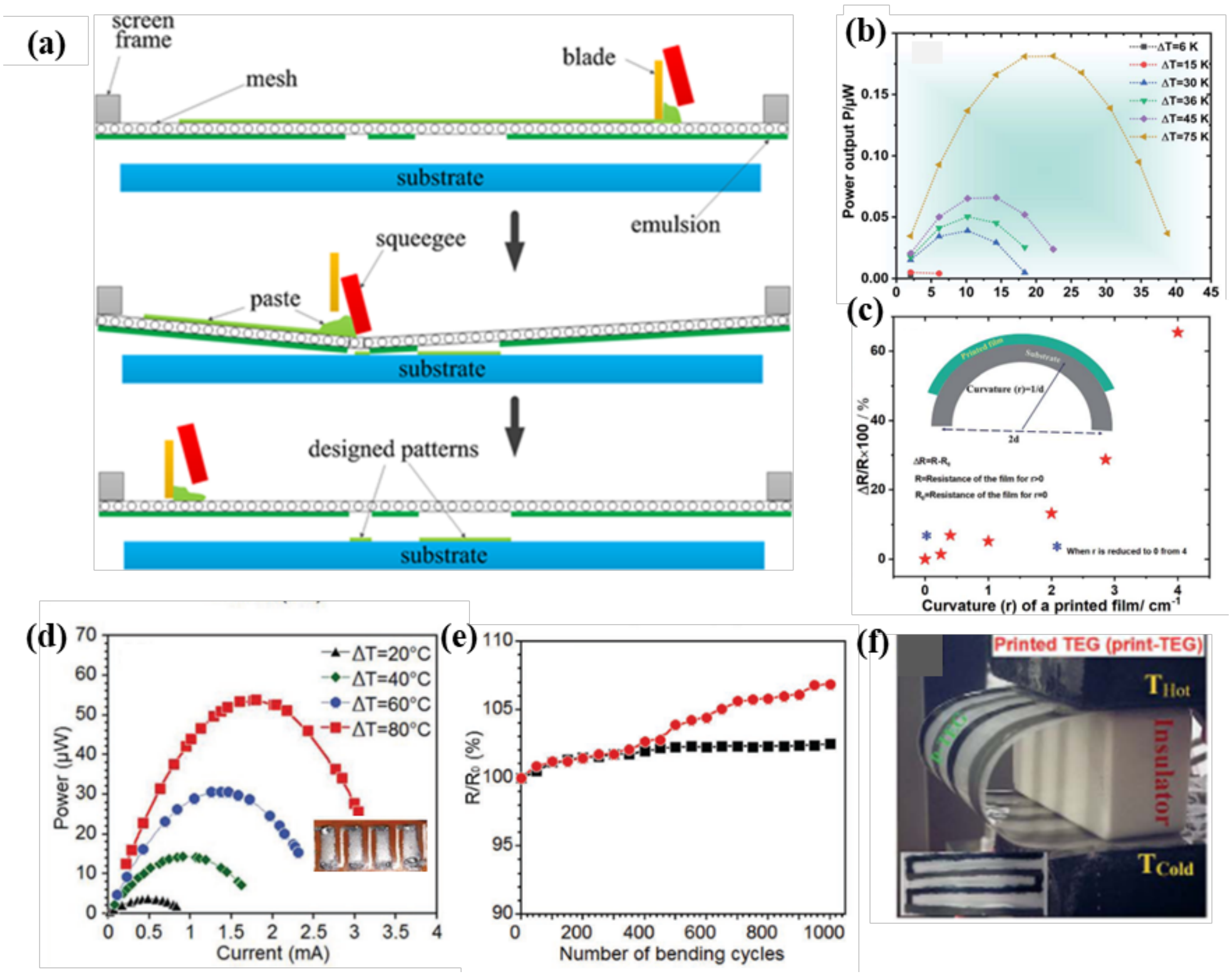

**Figure 8.** (a) Schematic diagram of screen printing. Screen printing work by Mallick, (b) power output, c) flexibility by resistance variation. Reproduced with Permission [80,81]. Copyright 2020, ACS Publications. and Varghese, d) Output Power, e) resistance variation with bending[82] f) flexible printed TEG by Mallick et al. Reproduced with Permission. Copyright 2020, The Authors. [80,81] Published by Wiley.

Due to the growing interest in harvesting waste heat into electricity for powering sensors and wearable devices, screen printing of TE materials is extensively investigated to fabricate flexible TEDs. Most of the works were focused on the popular BiTe-based n-type and p-type materials.[59] A ZT of 0.81 at room temperature was obtained by Shin et al. in 2017 by screen printing $Bi_2Te_{2.7}Se_{0.3}$ material on a fiberglass fabric substrate.[83] The screen-printing ink rheology was controlled by adding methyl cellulose binder. The printing thickness was limited between 10-700 microns. The printed sample was hot pressed to obtain high electrical conductivity as well as power factor. The removal of binder and solvent at high temperatures facilitated the high ZT of the material. Although they achieved high ZT for n-type materials, their p-type ZT was limited to 0.65 for $Bi_{0.5}Sb_{1.5}Te_3$ materials.

Mallick et al. demonstrated the highest ZT of n-type TE so far by screen printing silver selenide-based TE.[80,81] A power factor of 1700 μW/mK$^2$ for screen-printed n-type sample was achieved for $Ag_2Se$. Using this high-performance material, they produced a flexible TEG on a PEN substrate (**Figure 8**c,f) by coupling with commercial p-type PEDOT: PSS material, and a maximum power output of 0.19 μW was produced at a temperature difference of 75 K (**Figure 8**b).

In 2020 Varghese et al. showed the highest ZT of unity for p-type bismuth telluride-based flexible films using screen printing.[82] They used high energy ball milled p-type $Bi_{0.4}Sb_{1.6}Te_3$ powder with extra Te powder as a nanosolder to improve the sintering and enhance the charge carrier mobility of the printed materials. An ultrahigh power factor of 3 $mW/mK^2$ was achieved at room temperature for the $Bi_{0.4}Sb_{1.6}Te_3$-Te composite with average thickness around 27 microns. The printed films showed excellent flexibility by demonstrating only a 3-7% increase in resistance after bending for 1000 cycles (**Figure 8**d, e). Four $Bi_{0.4}Sb_{1.6}Te_3$-Te composite films were connected in series to make an in-plane TEG that produces a maximum power density of 18.8 $mW/cm^2$ at a temperature difference of 80 K.

### *3.1.2 Inkjet Printing*

Inkjet printing uses a continuous supply of ink droplets to fabricate film in a layer-by-layer method.[84] This printing technique has the advantage of high resolution and advanced control over the printing process. The continuous formation of a film or structure depends on the drying of the solvent as well. This method can be modified to supply drops of ink whenever necessary (Drops on Demand).[85] Despite the advantages of precise control over material deposition, the process is usually much slower compared to several other printing processes, making it applicable to thin TE films. The process of inkjet printing is depicted in **Figure 9**a. In a typical process, functional ink is deposited in a defined shape on a substrate.

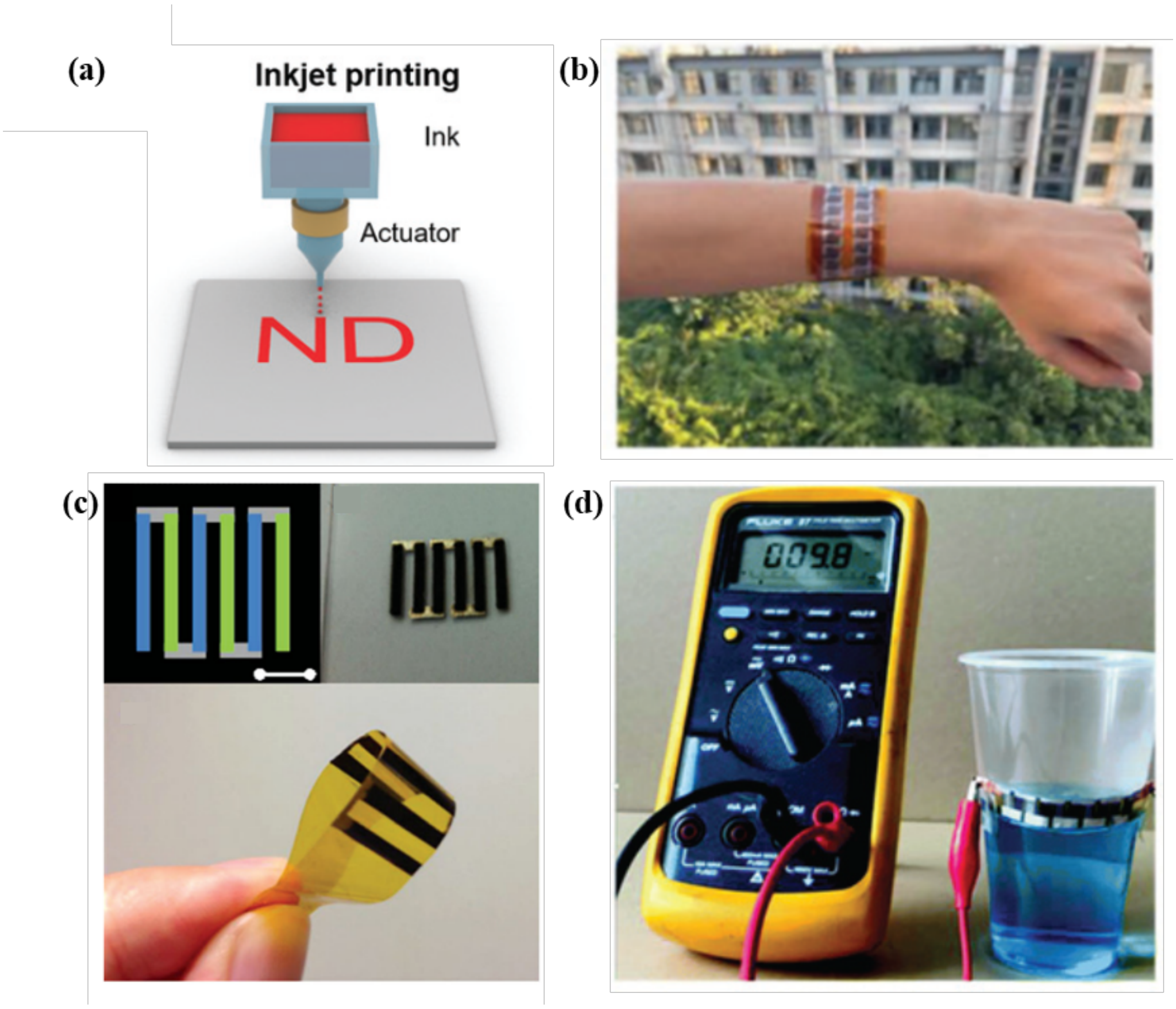

**Figure 9.** (a) Schematic of inkjet printing. Reproduced with permission. [28] Copyright 2019, RSC, (b) inkjet printed device by Du et al. Reproduced with permission [86], Copyright 2023, The Authors. Published by Wiley. (c) inkjet printed device by Lu et al.[87] Copyright 2014, The Authors. Published by Wiley. (d) power generation from a plastic cup using inkjet printed TE film by Chen et al. Reproduced with permission. [88] Copyright 2019, RSC.

There have been a few studies using inkjet printing for the fabrication of TEDs.[86,87] Lu et al. first used inkjet printing to make a power generation device using p-$Bi_{0.5}Sb_{1.5}Te_3$ and n-$Bi_2Te_{2.7}Se_{0.3}$ legs connected with printed Ag electrodes.[87] A maximum power factor of 183 $\mu W/mK^2$ in p-type and 77 $\mu W/mK^2$ in n-type legs was measured. A maximum power output of 0.341 mW was measured from the TED (**Figure 9**c). Since then, there has not been a notable improvement in the inkjet printed TEs. Recently, Du et al. used inkjet printing for printing nanowires of different chalcogenide-based TE ($Ag_2Te$, $Cu_7Te_4$, $Bi_2Te_{2.7}Se_{0.3}$). Their printed film showed a maximum power factor of 493.8 $\mu W/mK^2$ at 400K temperature. The flexible power-generating device (**Figure 9**b) based on the n-type $Ag_{2.1}Te$ could also survive one thousand bending cycles without notably deteriorating the electrical conductivity. A maximum output power of 101.3 nW was achieved with just 30 K temperature difference for an 800 nm thick inkjet printed device.[86] In 2018, Chen et al. inkjet printed $Bi_2Te_3$ and $Bi_{0.5}Sb_{1.5}Te_3$ nanowires and achieved the power factor of 0.18 mW/mK$^2$ and 0.11 mw/mK$^2$.[88] The TEG device was also flexible capable of going through at least 50 bending cycles. The in-plane device was capable of generating 9.8 mV from a glass of water with a temperature of 313 K (**Figure 9**d).

*3.1.3 Aerosol-jet Printing*

The aerosol jet printing (AJP) process can deposit a wide variety of conductive, semiconducting, and dielectric materials.[89] The AJP enables printing inks of a wide range of viscosities (1 cP to 1000 cP) with a superior spatial resolution (~10 μm) compared to other direct ink writing printing technologies.[90] This approach begins with aerosolizing inks using shear pressure or sonication force, forming a dense stream of aerosolized droplets. A sheath gas flow is applied to aerodynamically focus the aerosolized ink flow to realize high printing resolution. A unique feature of aerosol jet printing is its ability to perform 3D conformal printing on 3D curved components due to the relatively large stand-off distance between the nozzle and the substrate it can tolerate. A schematic showing the working process of AJP is presented in **Figure 10**a. [91]

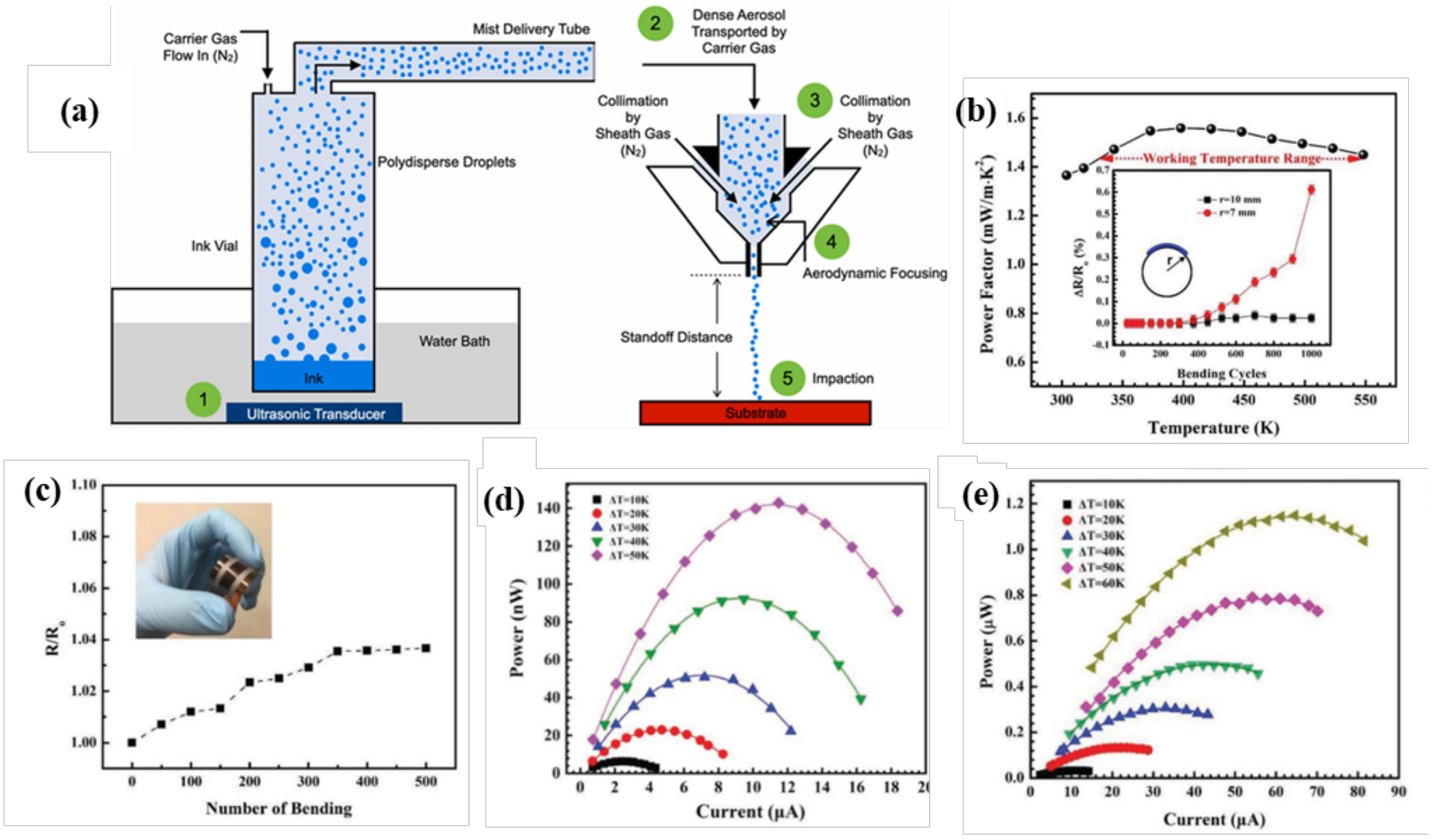

**Figure 10.** a) Schematic of aerosol jet printing process,[91] (b), (c) power factor and felxibility of aersol jet printed $Bi_2Te_{2.7}Se_{0.3}$ studied by Mortaza et al., Reproduced with Permission [92] Copyright 2019, The Authors. Published by Wiley. (d), (e) aerosol jet printing work by Courtney et al. Reproduced with Permission [93] Copyright 2020, The Authors. Published by Wiley.

There are few studies about printing TE materials using AJP.[94] Zhang's group developed flexible TEDs using aerosol jet printing of inorganic nanoparticle inks.[95,96] In 2019, Dun et al. printed 2D $Sb_2Te_3$ nanoplates combined with 1D Te nanorod, and achieved a power factor of 2.2 mW/mK$^2$ at 500 K. The flexible in-plane device they made delivered a power density of 7.65 mW/cm$^2$ with a temperature difference of 60 K.[97] Mortaza et al. studied aerosol jet printing of n-type $Bi_2Te_{2.7}Se_{0.3}$ material and obtained a power factor of 0.73 mW/mK$^2$ at room temperature.[92] The printed films were also flexible enough to survive 500 bending cycles without notable deterioration of the device resistance (**Figure 10**b, c). The printed device with four n-$Bi_2Te_{2.7}Se_{0.3}$ films joined in series could achieve a maximum power density of 2.7 mW/cm$^2$ at a temperature difference of 50 K. Courtney et al. printed $Bi_2Te_3$ nanocrystals using AJP on a polyimide substrate and obtained a power factor of 0.35 mW/mK$^2$ at 433 K.[93] The produced film was flexible and showed only a 23% decline in resistance after 1000 bending cycles (**Figure 10**d, e).

*3.1.4 Extrusion Printing*

Direct ink writing (DIW) using extrusion printing is a promising method for the utilization of 3D printing for manufacturing bulk TEDs.[59] This printing technique facilitates patterned deposition of the TE ink through a nozzle by several processes, such as screw-based extrusion system, pneumatic system, plunger system, etc. (**Figure 11**a). [98] The resolution of the printing is usually limited by the nozzle size and the ink rheology. Despite the limitation of lower resolution, the printing speed is usually higher than the other printing methods discussed before.[99] However, with a balance between resolution and printing speed, very complex shapes are possible to print, making this process a suitable approach for printing bulk TEDs.

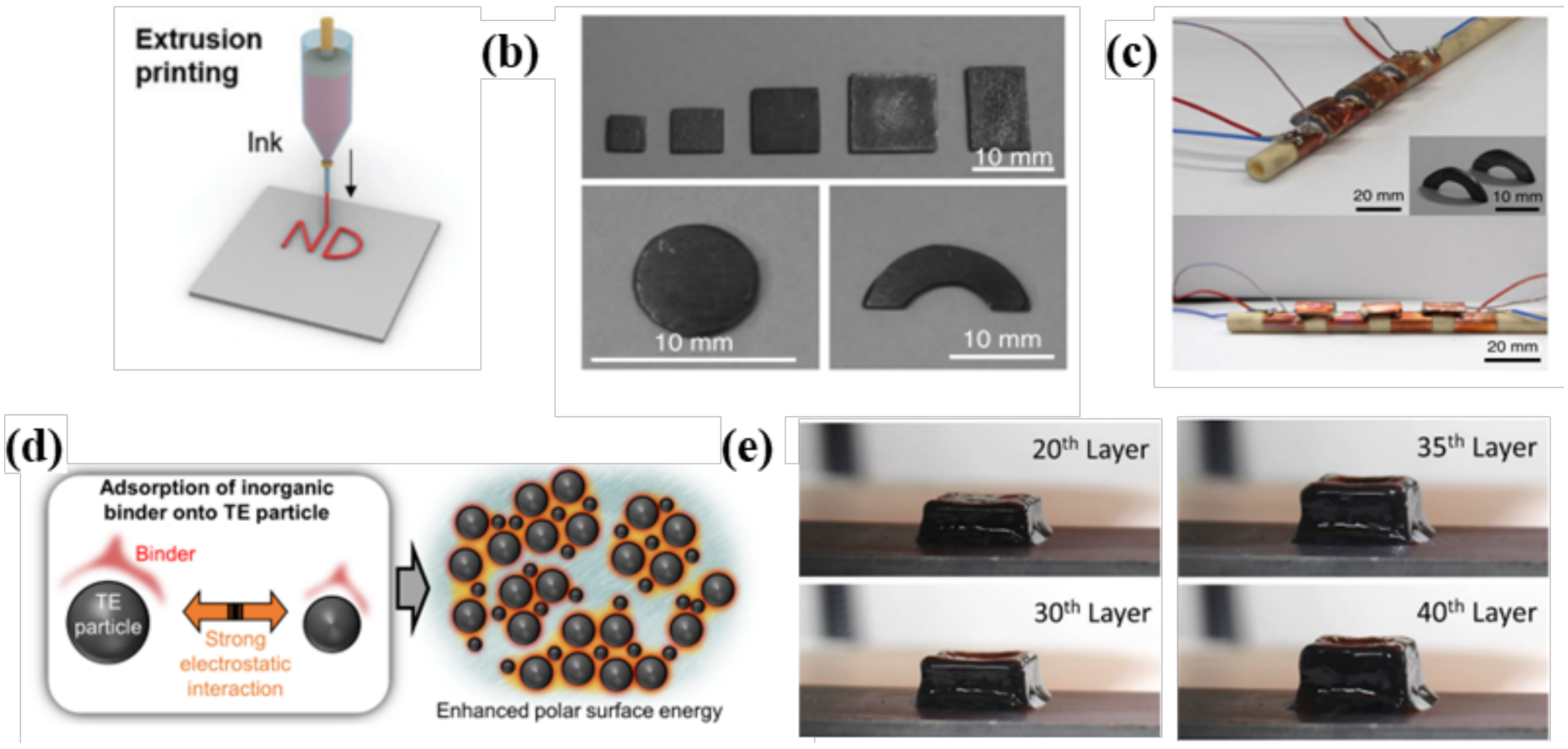

**Figure 11.** Extrusion printed TE materials (a) schematic of extrusion printing. Reproduced with permission. [28] Copyright 2022, RSC. (b) printed half-rings and other structures capable of producing energy from an (c) annular pipe. Reproduced with permission. [100] Copyright 2018, The Authors. Published by Springer Nature. (d) schematic of the effect of the inorganic ionic binder for a dense compacted sample and the associated 3D printed structure.[101] (e) 40 layers of TE ink made possible by utilizing the ionic binder. Reproduced with permission.[101] Copyright 2019, AIP Publishing LLC.

As a popular candidate for manufacturing bulk TEDs, several researchers have already printed different materials using extrusion printing. In 2018, Kim et al. reported *ZT* of 0.9 and 0.6 for p-type and n-type $Bi_2Te_3$-based extrusion printed materials.[100] They printed half rings conformable to a pipe to convert the waste heat energy of the pipe into electricity (**Figure 11**a). A TEG comprised of 3 pairs of TE half rings generated an output voltage of 27 mV and 1.62 mW electric power at a temperature difference of 39 K (**Figure 11**b, c). An inorganic binder was added to the ink to optimize the rheological properties of the TE ink and improve the densification of the particles for enhanced electrical properties. They later published a separate study regarding the process of controlling the rheology of the ink using the inorganic binder.[101] The inorganic ionic binder ($Sb_2Te_3$ chalcogenidometallate) enhanced the electrostatic interaction between the TE particles by modifying their surface energy, which enabled printing up to 40 layers without any collapse of the structure (**Figure 11**d, e).

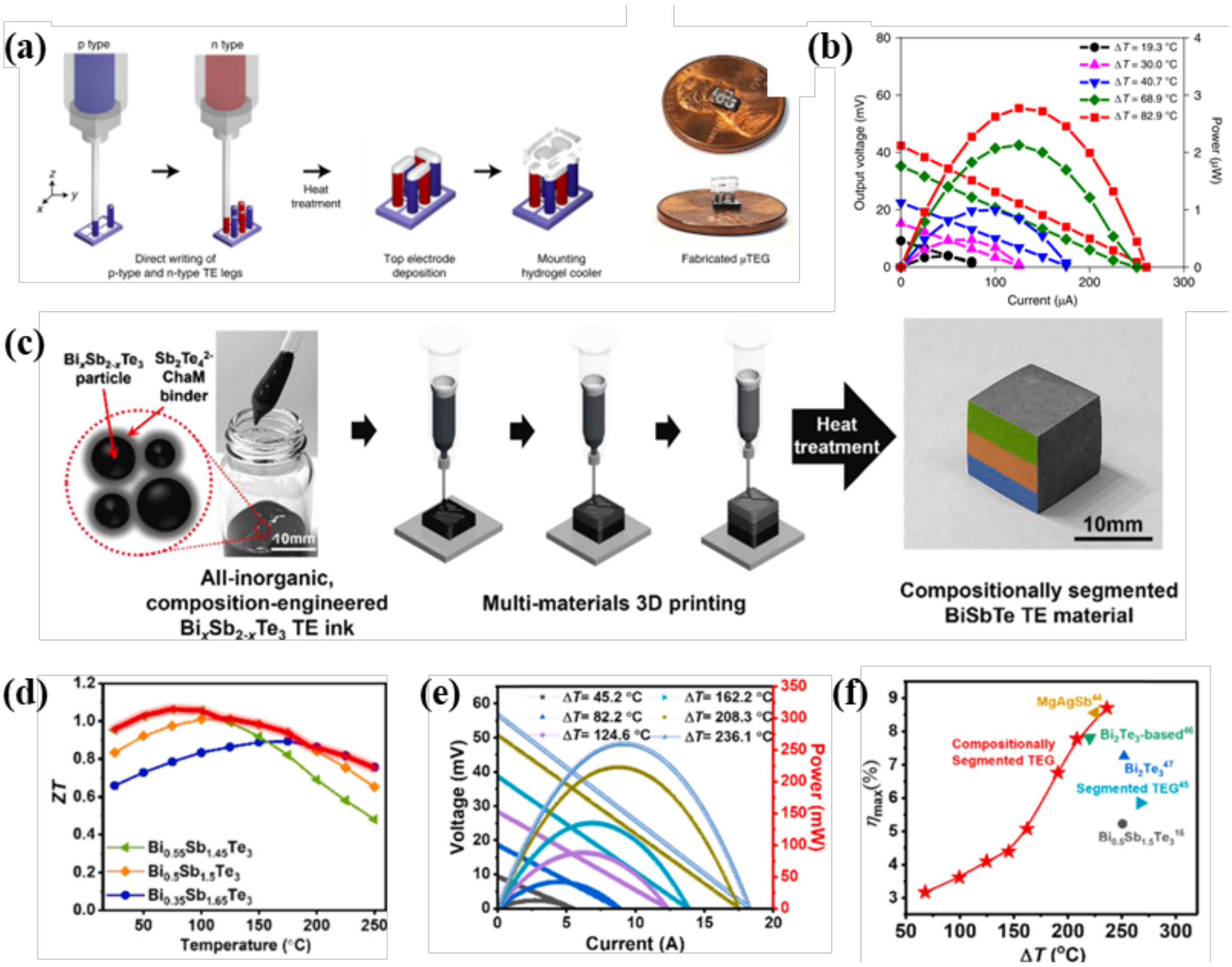

**Figure 12.** (a) TE microarchitecture and micro device by extrusion printing [102], (b) Output power and voltage generated by the μTEG, Reproduced with Permission. [102] Copyright 2021, The Authors. Published by Springer Nature. (c) Segmented TED preparation by extrusion printing [103], segmented TE (d) ZT, (e) output power and output voltage, (f) maximum efficiency and comparison to literature. Reproduced with permission. [103] Copyright 2021, Elsevier.

In 2021, Kim et al. showed complex microarchitectures capable of providing notable output power (**Figure 12**a, b).[102] The maximum *ZT* reached 1.0 for the p-type bismuth antimony telluride materials in this study. A μTEG was printed and produced a maximum output voltage of 42.4 mV and 2.8 μW of power at 82.9 K temperature difference (**Figure 12**b). Seong et al. printed a segmented TEG using $Bi_xSb_{2-x}Te_3$-based materials to achieve high *ZT* over a wide range of temperatures (**Figure 12**c, d).[103] They used three alloys of this material (x is 0.35, 0.5, and 0.55) to make a tri-segmented TEG, increasing their temperature range for high *ZT* from room temperature to 523 K (**Figure 12**d). The TEG showed an open circuit voltage of 56 mV at temperature difference of 509 K (**Figure 12**e). The maximum ZT of 0.94 in combination with the wide temperature range increased the efficiency to 8.7% (**Figure 12**f).

In summary, ink-based printing can be promising for low-cost manufacturing of TE materials and devices. TE properties of different printed inorganic materials using ink-based printing techniques are shown in **Table 2**.

**Table 2**. Thermoelectric properties of different printed inorganic materials

| Material | Printing Method | S (μV/K) | σ (S/cm) | $S^2\sigma$ (Temperature) (μW/mK$^2$) | κ (W/mK) | ZT | Ref. |
|---|---|---|---|---|---|---|---|
| $Bi_{0.5}Sb_{1.5}Te_3$ | InkJet Printing | 177 | ~24.50 | 77 (Room) | - | - | [87] |
| $Bi_{0.5}Sb_{1.5}Te_3$ | InkJet Printing | 83 | ~26.10 | 180 (Room) | - | 0.13 | [88] |
| $Ag_{1.9}Te$ | InkJet Printing | -80 | 200 | 493.8 (400 K) | | | [86] |
| $Bi_{0.5}Sb_{1.5}Te_3$ | Extrusion | 170 | ~400 | - (450 K) | 0.8 | 0.65 | [104] |
| $Bi_{0.4}Sb_{1.6}Te_3$ | Extrusion | 199 | ~53 | ~200 (473 K) | 0.5-0.63 | 0.9 | [100] |
| $Bi_2Te_{2.7}Se_{0.3}$ | Extrusion | -145 | ~ 85 | ~178 (473 K) | 0.5-0.63 | 0.6 | |
| $Bi_{0.55}Sb_{1.45}Te_3$ | Extrusion | 200 | ~70 | ~280 (350K) | 0.8 | 1 | [102] |
| $Bi_2Te_{2.7}Se_{0.3}$ | Extrusion | -120 | ~118 | ~170 (425 K) | 0.8 | 0.5 | |
| $Pb_{0.98}TeNa_{0.02}$ | Extrusion | 230 | 400 | ~2100 (700K) | 0.8 | 1.4 | [105] |
| $Bi_{0.5}Sb_{1.5}Te_3$ | Extrusion | 250 | - | - | - | 1.1, 373 K | [103] |
| $Cu_2Se$ | Extrusion | 186 | 310 | 630 (1000K) | 0.5 | 1.21 | [106] |
| $Bi_{0.4}Sb_{1.6}Te_3$ | Extrusion | 230 | - | - | - | 0.68 | [107] |
| $Bi_{0.5}Sb_{1.5}Te_3$ | Extrusion | 101 | 6.5 | 66 | 0.31 | 0.21 | [104] |
| $Bi_2Te_{2.7}Se_{0.3}$ | Extrusion | -113 | 3.9 | 49 | 0.31 | 0.16 | [59] |
| $Sb_2Te_3$ | Aerosol Jet Printing | 195 | 55 | 2200 (500 K) | - | - | [97] |
| $Bi_2Te_{2.7}Se_{0.3}$ | Aerosol Jet Printing | -163.4 | 272 | 730 (Room) | - | - | [92] |
| $Bi_2Te_3$ | Aerosol Jet Printing | - | - | 350 (433 K) | - | - | [93] |
| p-$Bi_{0.3}Sb_{1.7}Te_3$, n-$Bi_2Te_{2.7}Se_{0.3}$ | Screen Printing | - | - | - | - | - | [108] |
| $Bi_2Te_{2.8}Se_{0.2}$ | Screen Printing | -143 | 290 | ~590 (448 K) | 0.57 | 0.43 | [109] |
| PbTe-SrTe-2% Te | Screen Printing | 400 | 74 | 1380 (723 K) | 0.83 (Calculated) | 1 (723 K) | [110] |
| $Bi_2Te_3$ | Screen Printing | -120 | 240 | 363 (390 K) | ~0.57 | 0.26 | [111] |
| $Bi_{0.4}Sb_{1.6}Te_3$ | Screen Printing | 204 | 720 | ~3000 (Room) | 0.9 | 1 | [112] |

### 3.2 Rapid Processing and Sintering of Thermoelectric Materials

Post-printing processing and sintering are critical to consolidating the printed particles into a densified structure of desired properties. The large-scale production of TEG faces significant challenges due to the lack of high-throughput processing methods that can sinter TE particles rapidly while maintaining their high TE properties. Conventional thermal sintering in a furnace may take several hours to achieve desired microstructures and properties. The following sections will discuss the current progress on rapid processing for printed TE materials.

### *3.2.1 Intense Pulsed Light Sintering*

The Intense Pulsed Light Sintering (PLS) technique is a popular technique in sintering functional materials.[112,113] In this process, a Xenon flash lamp is used to produce high-intensity pulsed light, which can be absorbed by printed materials and converted into thermal energy to sinter the printed particles (**Figure 13**a).[92] The PLS is usually controlled by four different parameters – voltage, pulse duration, pulse delay, and pulse number. The PLS technique has the unique advantage of non-contact sintering of the materials, which makes it a highly scalable process. Typical sintering duration spans from milliseconds to several seconds, usually depending on the optical properties of the materials in use.

Several researchers have worked on the PLS of TE materials. In 2019, Zhang's group published a work on sintering n-type $Bi_2Te_{2.7}Se_{0.3}$ TE films produced by aerosol jet printing method.[92] The sintering parameters were optimized based on the TE performance measured. A maximum electrical conductivity of 270 S/cm could be achieved at an optimized pulse number. Increasing the pulse number beyond the optimal point increased porosity in the microstructure, which lowered the electrical conductivity (**Figure 12**b, c). The highest power factor was 730 µw/mK$^2$, which was among the highest for flexible n-type TEs. In 2022, they studied PLS on the n-type silver selenide material, and incorporated machine learning with expert intuition to optimize the sintering condition for films of different thicknesses.[114] With optimized sintering conditions, the power factor reached 2205 µW/mK$^2$ at room temperature. Moreover, the thin film could survive a thousand bending cycles without a notable decrease in electrical conductivity. A room temperature ZT of 1.1 could be achieved, which is still among the highest for flexible TEs. With the final optimized sintering condition, they fabricated a TEG consisting of six individual silver-selenide legs, which produced a maximum power density of 26.6 mW/cm$^2$ at a 70 K temperature difference. The flexible TEG was demonstrated as a wearable TEG as well capable of generating 1.4 mV at a 1.8 K difference of temperature between ambient and the human skin (**Figure 13**d).

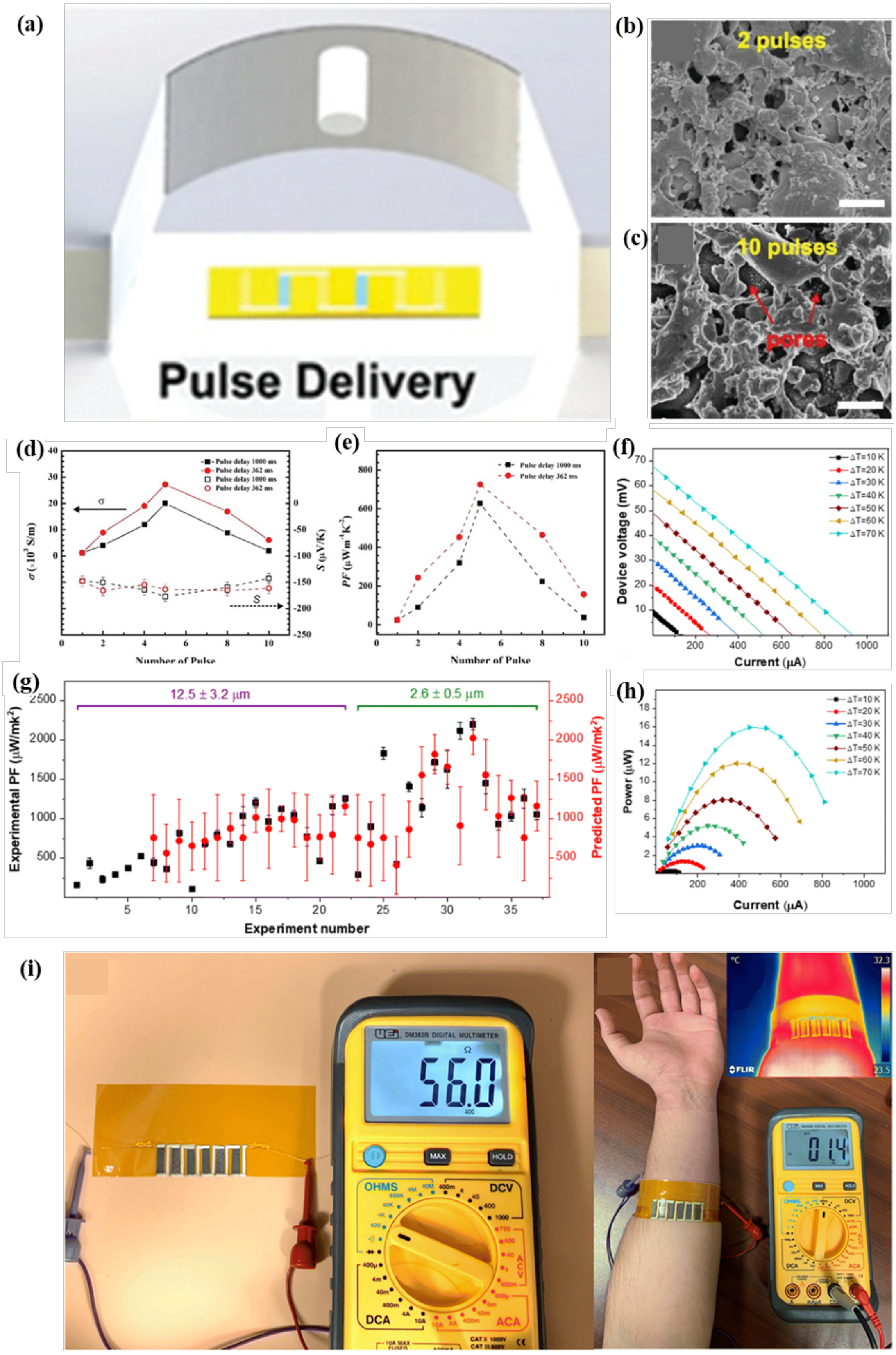

**Figure 13.** Intense pulsed light sintering work by Mortaza et al.[92,114] (a) pulsed light sintering schematic, (b), (c) effect of pulse number on porosity, (d), (e) optimization of power factor by optimizing pulse number, Copyright 2020, The Authors. Published by Wiley. (f)-(h) TED performance made by PLS procedure using machine learning-based optimization of sintering,

(i) TED capable of producing power at room temperature from human body heat. Reproduced with permission. Copyright 2022, RSC.

Mallick et al. worked with IPL-based sintering of p-type $Cu_2Se$ materials.[115] Later they worked with p-type $Bi_{0.5}Sb_{1.5}Te_3$ material in 2022.[116] Typically, this material requires high-energy sintering conditions due to the high melting point of the constituent materials. This study demonstrated a novel nano-soldering technique using a $Cu_2Se$-based inorganic binder to sinter the powders into a densified structure. Although the final flexible TE thin film (10-15 µm) contained porosity caused by the quenching effect and subsequent shrinking, the maximum ZT of 1.45 at 375 K for p-type and 0.765 for n-type at room temperature was achieved. A maximum power density of 5.1 $W/m^2$ at a temperature difference of 68 K was achieved by testing a half-millimeter thick (including the substrate) flexible TEG.

*3.2.2 Flash Spark Plasma Sintering*

Flash Spark plasma sintering (FSPS) is a novel concept recently used in sintering TE and ceramic materials.[117,118] As shown in **Figure 14**a, the FSPS process setup includes an electrical power source, a mechanical pressure device, and a vacuum chamber. In a typical process, the sample is usually pressed with a die while applying direct current through an electrode attached to the two ends of the die (**Figure 14**a). A vacuum chamber is usually used to avoid reaction with the atmosphere. The whole process usually requires less than minutes due to its high heating rate of up to $10^4$-$10^6$ K/min.[119] The required voltage is usually kept at a minimum (<10 V), while the applied current can be as high as 10000 A. For such a high current to pass through, the process is limited to highly conductive materials.

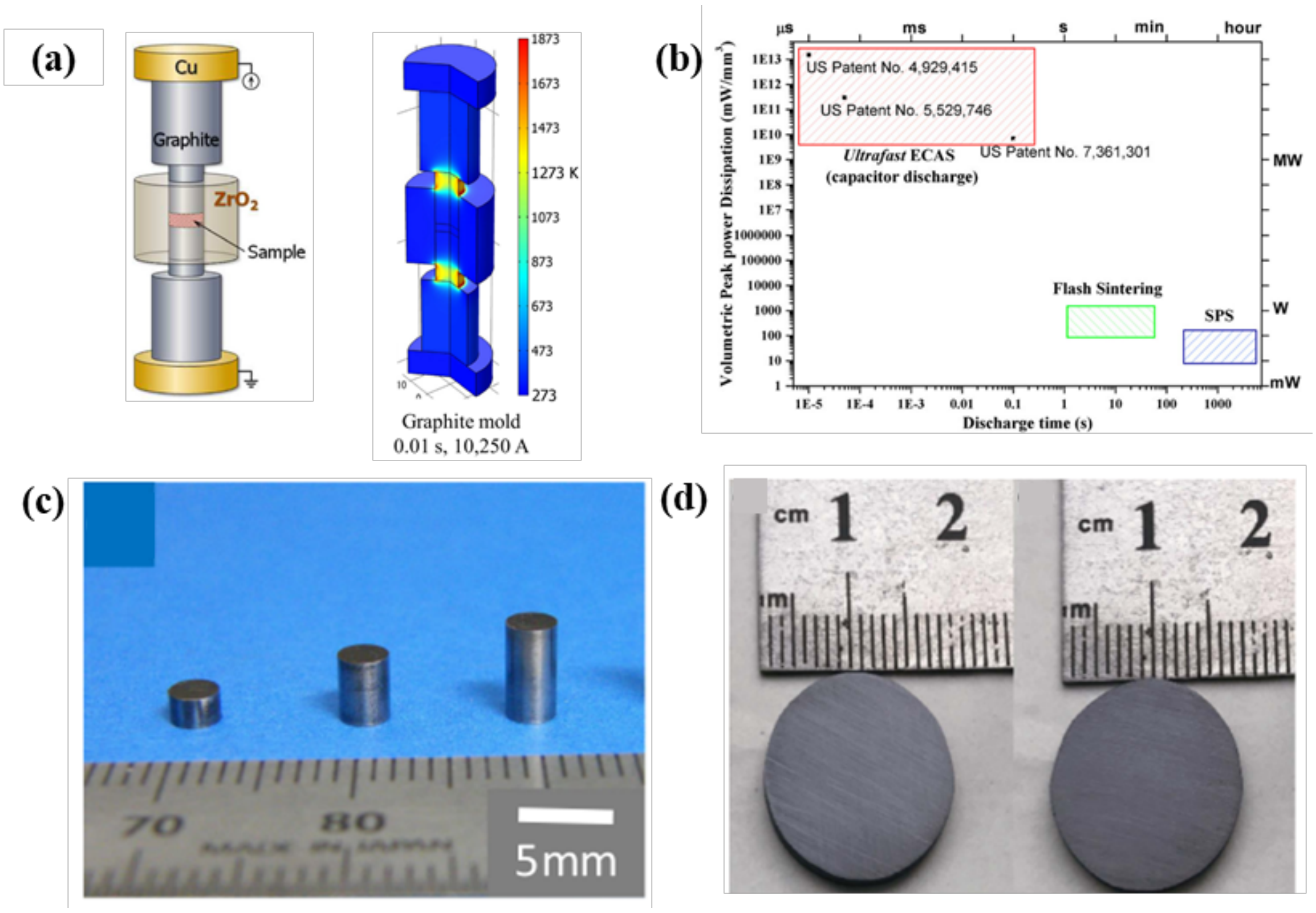

**Figure 14.** Flash spark plasma sintering of TE materials (a) schematic of the process and simulated temperature distribution. Reproduced with Permission.[120] Copyright 2022, Elsevier. (b) Comparison of FSPS and ECAS based on required time and power. Reproduced with Permission.[118] Copyright 2016, Talyor and Francis Publishing. FSPSed sample by (c) Mikami et al (Reproduced with permission. Copyright 2023, AIP Publishing.).[121] and (d) Min et al.[117] (Reproduced with permission. Copyright 2018, Elsevier.)

In 2017, Min et al. consolidated titanium suboxide ceramic powder and achieved a ZT of 0.085 at 1073 K.[117,118] In 2018. Mikami et al. applied FSPS to the conventional p-type $Sb_2Te_3$ material.[121] With a current feed duration of just 1s, bulk samples of thickness higher than mm could be sintered with less porosity compared to the bulk sample prepared by furnace sintering. Seebeck, electrical conductivity, and thermal conductivity were measured. Because of the difference in the measurement plane of electrical and thermal conductivity, they could not estimate the ZT based on their result. While the Seebeck coefficient was similar, the electrical conductivity was substantially lower in the FSPSed sample than the furnace-sintered sample. Moreover, the FSPSed sample was highly oriented, which might also cause lower electrical conductivity. They later published another research on a non-conventional TE ($Fe_2$Val) with higher material density but low ZT.[122] In 2019, Hao et al. synthesized and sintered $Cu_{1.97}S$ bulk sample within only 30 seconds utilizing pulsed electric current assisted FSPS.[123] The final sample was dense and heterogeneous in the microstructure. They could achieve a ZT of 0.72 at 873 K of temperature. The combination of synthesis and sintering can open new doors for chemical reacting and making conventional TEs using this method. In 2022, Mikami et al. investigated the FSPS again with $Bi_2Te_3$ materials.[120] They showed that the p-type and n-type conduction of the TE can be easily controlled by changing the amount of current supplied through the sample. Complete n-type transformation of $Bi_2Te_3$ could be achieved at high current density. They also fabricated Se-doped n-type $Bi_2Te_3$, which showed comparable TE performance to the furnace sintered sample.

Although the rapid solidification technique using current is primeval, the process requires more attention from the TE community to be applicable in manufacturing. The change of microstructure and the transport properties by the electron flow needs to be widely investigated for a better understanding of the process and property. Overall, there have been very few studies on the rapid manufacturing part for TE materials due to the more obvious obstacle of other processing problems, such as obtaining high average ZT and efficiency, and practical issues like electrical contact processing.

### 3.3 Electric Contact Processing

Ensuring a solid electrical connection between TE elements is a key requirement of efficient TE power generation and cooling devices. TE elements are usually connected by a conductive metal electrode. The connected thermoelements are coupled with a load resistance or a surface to use for power generation or cooling/heating purposes, respectively. The selection of the electrode depends on the specific material of use. A TED usually requires high thermal conductivity and electrical conductivity of the electrode. The electrode material is either directly connected to the TE element or via different barrier layers to protect the thermoelement from chemical reactions with the electrode. The resistance between the electrode and the TE materials also affects the final performance of the TED. Joshi et al determined the relation of electrical contact resistance with ZT, output power, and the efficiency of the TEDs by following equations[124,125]:

$$\eta = \frac{\frac{T_H - T_C}{T_H}}{\{\left(1+\frac{2rd_C}{l}\right)^2 [2-0.5\left(\frac{T_H-T_C}{T_H}\right)+(\frac{4}{ZT_H})(\frac{1+n}{1+2rd_C})]\}} \quad (2)$$

$$\langle ZT\rangle_D = \frac{L}{L+2R_C\sigma}\langle ZT\rangle_m \quad (3)$$

$$P = \frac{S^2}{2\rho}\frac{AN(T_H-T_C)^2}{(n+1)\left(1+\frac{2r\,d_C}{l}\right)^2} \quad (4)$$

where $N$ is the number of TE pairs assembled in a single device, $d_c$ is the thickness of the contact layer, and $T_H$, $T_C$, $A$, and $l$, are the hot- and cold-side temperatures, the cross-sectional area and length of the TE element, respectively. From the equations, this is evident that translating the material $ZT$ into device $ZT_D$ and lowering the electrical contact resistance is mandatory. A $ZT$ near 2.6 for SnSe has already been published. However, due to a lack of research on the device fabrication electrical contact resistance has not yet been optimized for different materials postponing the progress of the TEDs.

There is extensive research already in the literature for lowering the electrical contact resistance of TEDs made of bulk thermoelements.[124,126–130] Since we are more focused on printed TE in this review paper, we will discuss the current progress on enhancing the electrical contact of electrodes and thermoelements fabricated by printing processes. There has not been an exhaustive search for finding the most suitable joining methods for printed TEs. Kim et al developed a novel post-processing technique to lower the contact resistivity of p-type and n-type BiTe film to electrode by 16-fold and 3-fold, respectively. They used a method called the reduction ambient annealing process, which could remove the surface oxides of the TE film efficiently. The novel process could also modulate the Te distribution along the surface to enhance TE performance as well. The RAA process was followed by screen printing a bilayer of Ni/Ag paste to reduce the chance of diffusion of electrode materials. Finally, HF-assisted cleaning could provide the lowest possible contact resistivity. The said process could enhance the final ZT from 0.301 to 0.595 and 0.247 to 0.481 for the p-type and n-type leg, respectively.[131] In another work by Joshi et al., pulsed light annealing is performed to enhance the electrical conductivity of the contact fabricated by electroplating. The Ni barrier layer was annealed with a high-intensity Xenon flash lamp. Due to the smaller grain size of the annealed barrier layer, the contact resistivity for both the p-type and n-type layers could be lowered below 10 $\mu\Omega/cm^2$.[132,133] Although this process uses a hot pressed TE sample, the process can be applicable for the printed TE as well.

## 4. Applications of Thermoelectric Devices

The foundational concepts of the TE effect, particularly the conversion of heat into electrical energy (Seebeck effect) and its reverse (Peltier effect), have driven notable progress in TED research. To date, TED has mainly been widely used in waste heat recovery power generation (capture and reuse of waste heat from a variety of sources, including automotive, industrial, human, solar, and nuclear) and cooling equipment. This chapter will take an in-depth look at these two broad categories of TED applications.

### 4.1 Waste Heat Recovery

Waste heat refers to the excess thermal energy generated as a secondary outcome in energy systems. TEGs are engineered to capture this leftover thermal energy and convert it into electrical power. Unlike conventional heat engines that use a working fluid to turn thermal energy into kinetic energy, TEGs use charge carriers as their "working fluid." This enables a direct and motionless transformation of heat into electricity.[28] In recent years, TEG systems have become an attractive alternative or complement to power systems in a variety of applications, including automotive engines,[133–135] industrial electronics,[136,137] self-powered wireless platforms,[138,139] health tracking and monitoring systems,[140,141] and aerospace applications[142,143] due to their flexibility, cost-effectiveness, and reliability. The following sections detail the TEGs used in these different applications.

#### *4.1.1. Automotive Waste Heat Recovery*

The energy use patterns of cars indicate that over two-thirds of fuel energy becomes wasted heat, as the efficiency of most combustion engines doesn't exceed 30%. [144,145] Therefore, employing TEGs to capture and convert this waste heat into electricity is a key strategy for improving overall energy efficiency. Exhaust gas temperatures in cars, post-combustion, can reach as high as 1073 K, and even in downstream areas, they remain over 500 K.[146,147] This provides a heat source for TEGs in the mid to high-temperature range. Major car companies worldwide, such as BMW,[146] Ford,[148] Renault,[147] and Honda,[149] are investing in exhaust heat recovery technology, developing systems that use TEGs. In addition to the efforts of industry, the academic community has also made significant academic contributions in this field. For example, Zhang et al. showcased a powerful 1 kW TE system designed for automotive waste heat recovery in **Figure 15**a.[150] They constructed devices using nanostructured TE materials that achieved a power density of 5.26 $W/cm^2$ with a temperature differential of 500 K. There are also some studies focusing on TEGs for low-temperature automotive waste heat recovery, designed to operate with a hot-side temperature under 333K.[151,152]

Due to the broad range of temperatures involved in internal combustion engines, the efficiency of energy conversion can be significantly enhanced by assembling various types of TE elements. Segmented TEG designs are frequently employed to take full advantage of the large temperature difference.[146,153,154] LaGrandeur and his team engineered a tri-stage configuration of segmented TE substances: N- and P-type $Bi_2Te_3$ for low-temperature spans (< 523 K), P-TAGS and N-PbTe for intermediate temperatures (523–773 K), and skutterudite (SKD) materials (P-$CeFe_3RuSb_{12}$ and N-$CoSb_3$) for high-temperature spans (773–973 K).[146] The team also proposed an innovative design for these segmented materials, permitting manipulation of the thickness of TE elements and their thermal expansion coefficients to enhance conversion efficiency. In addition to experimental methods, the performance of automotive TEG systems has been modeled using a fluid-thermoelectric multiphysical numerical model, taking into account factors such as mass flow and exhaust temperature at different vehicle speeds.[155] The modeling helps improve the peak power output and conversion efficiency of the TEG system while also understanding how the location of the TEG affects the output performance. Zhang and colleagues presented a 3D model focused on a high-temperature (>873 K) automotive TEG system. Their research emphasized fine-tuning heat exchanger design variables and the TEG design to maximize fuel efficiency.[156] Using 72 nanostructured bulk half-Heusler TE modules in the optimized TEG, they observed a fuel efficiency boost of 2.5% with a SiC heat exchanger and 2.0% with a SS 444 heat exchanger. Although TEGs offer durability, longevity, and high-temperature stability, making them suitable for cars. However, they face challenges such as high material prices, the need for thermal management, and low conversion efficiency. Ongoing research aims to address these issues to improve the performance/cost ratio of TEG for automotive waste heat recovery.

#### *4.1.2. Industrial Waste Heat Recovery*

Industrial waste heat can also be captured using TEGs to improve energy efficiency across industries. The recovered electricity can be used to operate equipment or machinery within the same industrial facility, thereby reducing the overall dependence on external energy sources. Heat pipes, a frequently encountered form of industrial waste heat, are passive heat transfer tools often used to manage disparities between supply and demand. Tang and colleagues introduced a unique method for recovering waste heat - the Heat Pipe Thermoelectric Generator (HPTEG). This system combines a potassium heat pipe and an SKD-based TEG to recover waste heat and simultaneously produce electricity.[157] These potassium heat pipes showed a high thermal efficiency of 95% and maintained superb isothermal conditions with a temperature differential within 288 K. The TEG achieved a 7.5%

TE conversion efficiency, while the overall HPTEG system hit 6.2%. The TEG achieved a peak power output of 183.2 W, with an open-circuit voltage of 42.2 V when its hot surface temperature was at 898 K. Additionally, Madan et al. invented a flexible planar TEG capable of extracting heat from industrial pipes to power a wireless sensor network. [158] They used screen printing on pliable substrates to create 420 Ag/Ni thermocouples, making the device adaptable to the pipe's cylindrical form. With a 127 K temperature differential, this TEG generated 308 mW of power.

Aiming to enhance power generation efficiencies, Zhang et al. explored a micro-combined heat and power (micro-CHP) system.[159] This configuration incorporated nanostructured bulk high-temperature TEGs into a gas-powered combination boiler intended for space heating and providing domestic hot water. The TEGs, composed of highly efficient nanostructured bulk half-Heusler alloys, exhibited a maximum ZT value of 0.9 at 700°C (p-type) and 1.0 at 500°C (n-type). The result of this setup was a remarkably elevated power density of 2.1 $W/cm^2$ and an electrical efficiency of 5.3% when exposed to a 500 K temperature difference between its hot and cold sides. This system capitalizes on the latent exergy existing between combustion gas and water, converting thermal energy into electrical power with an efficiency of 4%.

The heat from nuclear waste has often been overlooked, yet it contains vast amounts of energy. This energy holds significant potential to be harnessed by TEGs. Kempf and his team evaluated a TEG made from nanostructured bulk half-Heusler (HH) materials under the extreme radiation environment of a reactor core. [160] Even after 30 days in the MIT Nuclear Research Reactor, this TEG consistently produced an electrical power density exceeding 1140 $W/m^2$ despite being exposed to an intense fast-neutron fluence of 1.5 × 1020 $n/cm^2$. As depicted in **Figure 15**b, the coolant flow and ample radiation in the nuclear reactor highlight the TEG as a prime technology for power harvesting within the reactor core.

*4.1.3. Human Body Waste Heat Recovery*

Using TEGs to convert the body's waste heat into energy is a promising concept, especially for wearable technology. The consistent metabolic heat can potentially power devices such as health monitors and small medical instruments.[161] Many of these wearables require minimal energy, even as low as sub-milliwatts,[162] which TEGs can achieve. Furthermore, incorporating TEGs in devices like pacemakers could reduce the need for battery charges, enhancing patient safety and comfort.[163] Researchers have explored various body locations for TEG placement, leveraging the temperature difference between the environment and skin, including areas like the wrist, chest, and even clothing items.[163–165] However, due to the low ZT value of TE materials suitable for biomedical use[166] and the slight temperature gradient of the active layer of TEG devices, there are still some obstacles to designing wearable TEG.[167] Moreover, the high thermal resistance between the skin and the hot side of these modules further complicates the matter.[168] These limitations affect the power efficiency of these wearable TEGs.

To overcome the challenges of inadequate flexibility and excessive weight in wearable technology, a technique has been proposed for creating wearable organic TEGs using printed lightweight carbon nanotubes.[169] The scientists demonstrated n-type and p-type films that preserved superior flexibility and satisfactory chemical stability even at 300°C in air. A noteworthy application of these wearable and flexible TEGs is a miniaturized accelerometer that utilizes body heat to generate power. This TEG made up of 52 pairs of cube-like TE legs on a flexible printed circuit board, can produce an open-circuit voltage of 37.2 mV with a temperature difference of 50 K. This output is adequate for operating a 3-axis mini accelerometer to detect body movement (as shown in **Figure 15**c).[170] Substantial progress has also been made in increasing the power density of TEG wearables.[93,141,164,171–173] The development of efficient TE materials and advanced manufacturing technologies has greatly expanded the possibilities for wearable device applications, using micro-TEGs as power sources.[174,175] As depicted in **Figure 15**d, a TEG medical device powered by body heat and a wireless pulse oximeter have been showcased.[163] Furthermore, Hyland et al. developed an optimized TEG with heat spreaders on both sides, featuring a compact design suitable for wearable applications.[176] The power generated on various parts of human skin was measured and compared, identifying the upper arm as the ideal location for powering electrocardiogram (ECG) sensors.

In the context of implantable TEGs, ensuring that the materials or coatings used on the device are biocompatible and non-toxic is crucial. A thermally conductive, biocompatible membrane has been applied to the device's surface to improve biocompatibility.[162] Subsequently, the same team developed biocompatible TEGs successfully implanted into rabbits.[177] A consistent temperature difference of approximately 0.5 K was recorded, and both TEGs could generate a voltage of over 20 mV. The energy produced by these TEGs was sufficient to power a clock circuit with higher energy demands than a standard cardiac pacemaker. This successful *in vivo* experiment demonstrated the potential for TEGs to harvest energy in medical applications, offering valuable insights for future biomedical applications.

Conductive polymers, widely used in biomedical devices, have seen significant strides in enhancing their conductivity and power factors for various wearable applications in recent years.[178–183] The inherent properties of these polymers, including their low thermal conductivity, lightweight, high flexibility, and excellent mechanical compliance, make them ideal for use in the fabrication of TEGs through assembly, printing, or coating methods.[184,185] The polymer PEDOT: PSS, in particular, has been extensively investigated as a promising TE material.[186] With its exceptional electrical conductivity, low thermal conductivity, easy processability, and decent ZT,[187] it and its composites are suitable for flexible TEGs in biomedical uses. By adding carbon nanotubes or tellurium nanowires, the TE efficiency of these polymer-based TEGs can be enhanced further.[166] To introduce properties like stretchability, other elastic polymers can be integrated with the conductive polymer framework.[188] With the combination of innovative material exploration and carefully designed architectures,[167,189] we can anticipate substantial progress in implantable/wearable TEGs in the next few decades.

#### *4.1.4. Solar Thermoelectric Generators*

Solar Thermoelectric Generators (STEG) are TE instruments that harness energy from sunlight or solar radiation.[190,191] There are already many examples of STHG, such as the advanced flat panel STEG developed by Kraemer et al., which achieves an efficiency of 4.6%,[191] and the printed STEG with 10 pairs of p-n TE legs that produce an open circuit voltage of 55 mV and an output power of 4.44 mW (**Figure 15**e).[192] While efficiency improvements are necessary to enhance the practical applicability of STEGs.[193] In the case of STEGs, similar to photovoltaic (PV) cells, the enhanced power conversion efficiency critically depends on effective light absorption. Implementing efficient solar heat collectors and solar concentrators with wavelength-selective capabilities is fundamental for enhancing device performance. In addition, device structure optimization can further augment STEG performance. For instance, Xiao and his team introduced multi-stage configurations using bismuth telluride and filled-SKD materials. They evaluated the TE output of various designs, including single-stage, two-stage, and three-stage unicouples. The findings indicated that the TEG conversion efficiency rose with an increase in the number of stages.[194] Furthermore, employing computer simulations and modeling can help improve STEG efficiency.[195]

Even though advancements have been made in enhancing solar cell efficiency, they still produce a substantial amount of waste heat that isn't harnessed in the PV process. TEGs, when integrated with PV cells, can serve as a supplementary system that transforms this waste heat into useful power, enhancing the overall efficiency of hybrid energy systems.[196–198] Milijkovic and team studied a hybrid solar TE (HSTE) system, which employs a thermosyphon for passive heat transfer to a bottoming cycle.[199] This setup employs parabolic trough mirrors to concentrate solar energy on a particular surface layered with TE material. The results indicated that an HSTE system can achieve maximum efficiency of 52.6% when the solar concentration is at 100 suns and the bottoming cycle temperature reaches 500°C. Beeri et al. proposed a hybrid PV/TEG system that combines concentrated sunlight conversion via PV and TE methods with a concentration factor (X) up to ~300.[200] The system incorporates a TEG and a multi-junction PV cell. Up to X≤200, the system's peak efficiency of ~32% was largely attributed to the PV cell. As X and system temperature increased, the efficiency contribution transitioned from the PV cell to the TEG, which achieved ~20% at X≈290. Notably, under high solar concentrations, the TEG's cooling impact was observed to outweigh its direct electrical input, leading to a cumulative efficiency contribution of roughly 40% at X≈200. This introduces a new optimization strategy for the system, considering the temperature sensitivity of the PV cell and the balance between the power generation of the TEG and its cooling capability. The study suggests that with more advanced PV cells and TE materials, this hybrid system might surpass a total efficiency of 50%.

*4.1.5 Radioisotope Thermoelectric Power Generation*

Radioisotope Thermoelectric Generators (RTGs) are power generation systems that transform the heat produced by radioactive decay into electricity. Thanks to their high energy density and minimal maintenance requirements, RTGs are frequently used in harsh environments like the deep ocean or outer space.[201]

To improve RTGs' efficiency, Holgate and his team introduced an improved version of the Multi-Mission Radioisotope Thermoelectric Generator (eMMRTG).[202] This proposed model includes 768 SKD-based thermocouples with nickel-based connectors. Operating under high temperatures of 600–625 °C and low temperatures of 100–200 °C, the SKD-based eMMRTG generated an initial electric power of 90–105 W and attained a conversion efficiency of 7.6–8.3%, which marks a substantial enhancement over the 6% efficiency of the PbTe/TAGS MMRTG. Moreover, considerable research has been directed toward creating low-power RTGs to power low-energy devices in space systems. For example, Liu and his team designed a micro-radial, milliwatt-level RTG featuring a radioisotope heat source and four TE modules encased in an aluminum cylinder. [142] Using an electrically heated aluminum oxide-based helm with an internal resistance of 3.2 Ω, they replicated an Am-241 isotope heat source. This heat source, with dimensions of 7 × 7 × 27 $mm^3$, was housed in a copper-based shield. The TEGs were crafted from low-temperature $Bi_2Te_3$. When the heat source power is 0.1W, the RTG output voltage is 92.72mV, and the electric power is 149μW.

The advent of additive manufacturing has introduced new technologies, such as screen printing, enabling rapid prototyping of a wide range of compact RTGs with fine geometry, thus expanding the range of their potential applications and manufacturing methods.[143,203,204] Yuan et al. fabricated a device (as shown in **Figure 15**f) that, when combined with a 1.5 W isotope heat source, generated a short-circuit current of 0.329 mA and an open-circuit voltage of 68.4 mV. It achieved its peak output power of 5.8 mW at 39.2 mV. [143] While research on printed RTGs remains somewhat limited, it is anticipated that the customizable and sophisticated TE modules enabled by printing techniques, combined with printable high-ZT materials, will enhance the power density and efficiency of RTGs, making them more suitable for use in extreme conditions.

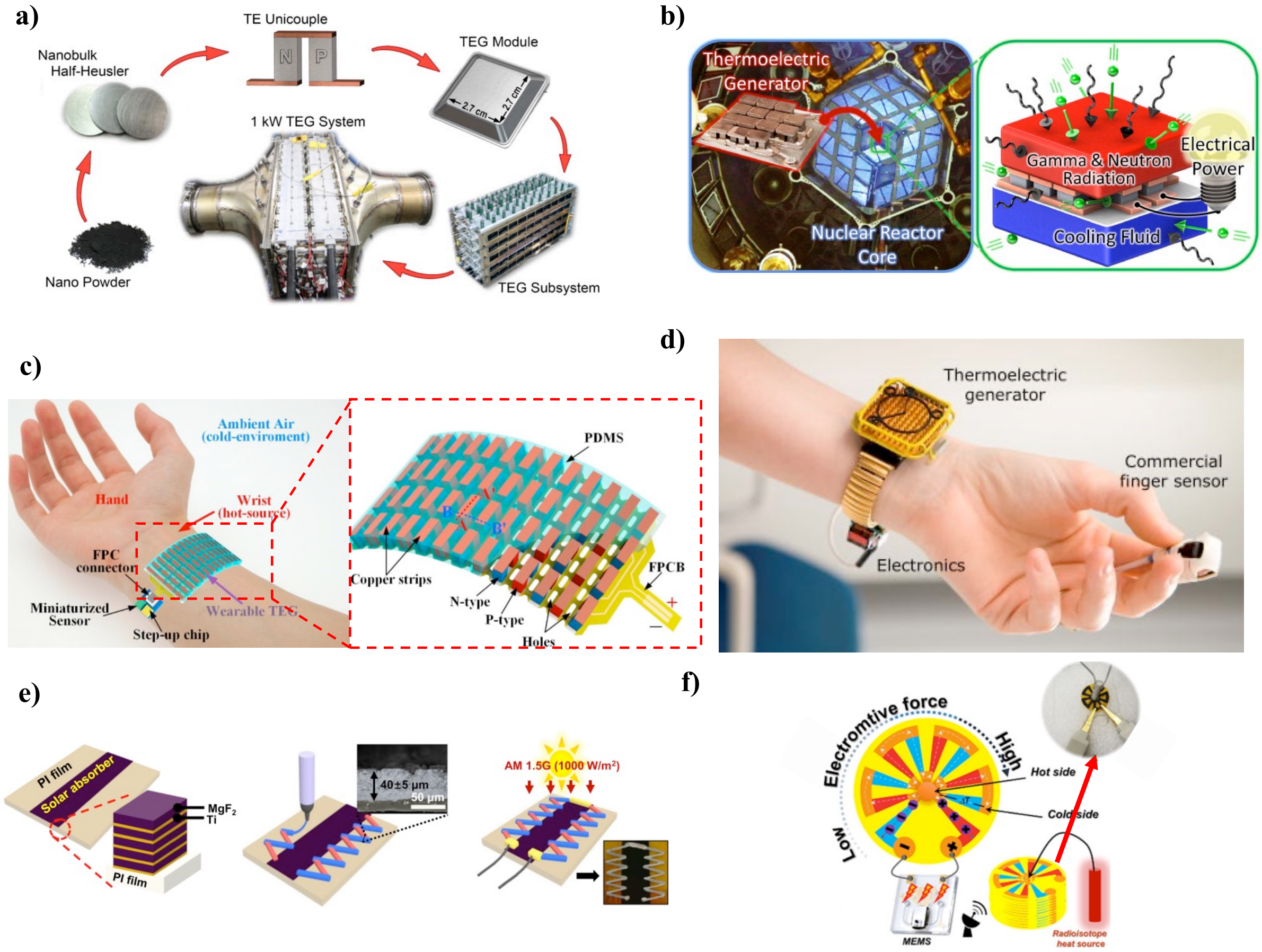

**Figure 15.** Examples of TE-based waste heat recovery devices. (a) Pictures illustrating the fabrication processes of the high-performance TEG. Reproduced with Permission.[150] Copyright 2015, The Authors. Elsevier. (b) A photo of the MIT nuclear reactor laboratory's

core (left) has an inset of the nanostructured half-Heusler TEG. On the right, a schematic illustrates a TEG generating electricity from heat via a gamma susceptor and cooled by a fluid on its cold side. Reproduced with Permission.[160] Copyright 2022, The Authors. Elsevier. (c) Schematic representation of the wearable TEG positioned on the wrist to energize a motion tracking sensor, along with the structural layout of the TEG. Reproduced with Permission.[170] Copyright 2018, Elsevier. (d) A body-powered wireless oximeter. Reproduced with Permission.[163] Copyright 2009, AIP Publishing. (e) Illustration of the printing method and operational concept of a printed STEG exposed to solar radiation, featuring a close-up of the TE legs. Reproduced with Permission.[192] Copyright 2017, Elsevier. (f) Schematic of MRTG with the inset of the sample of MRTG. Reproduced with Permission.[143] Copyright 2018, Elsevier.

### 4.2 Thermoelectric Cooling Device

The significance of TE cooling lies in its ability to enable precise temperature control in a solid-state form, eliminating the need for traditional refrigerants and compressors. Materials play a pivotal role in determining the efficiency of TE modules, and advancements in TE figure of merit ZT have led to improved cooling performance.[4,14,205–207] While the energy generation via TE materials is still not mature enough to revolutionize waste heat recovery, TE cooling can be highly effective due to its less harsh working environment (temperature difference less than 100 K with near ambient operating temperature). Different types of cooling applications of TE materials are shown in **Figure 16**. We will discuss several applications in detail.

Car climate control through a TE temperature controller is an already commercialized concept.[208] There have been investigations for localized car climate control (**Figure 16**a) as well as the development of automated car safety seat cooling systems using TE coolers.[209–211] These systems aim to mitigate the risks associated with high temperatures inside parked vehicles, especially for children and individuals with physical challenges. Traditional measures like tinted windows and circulation fans are insufficient in reducing peak temperatures to safe levels. TE cooling offers a potential solution to address this issue by maintaining a comfortable temperature within the car seat, ensuring passenger safety and well-being.[209] Moreover, TE refrigeration is a promising concept for replacing commercial vapor compression-based refrigeration systems. However, the current TEDs still need much improvement in terms of efficiency and cost reduction to match the conventional refrigeration system.[212–214][215–217] Therefore, TE cooling devices are still limited to niche applications.

The integration of TE systems with photovoltaic panels, a concept explored in depth, holds the potential to create solar TE systems capable of harnessing sunlight to power cooling processes. This integration not only enhances energy efficiency but also aligns with sustainable practices.[218] **Figure 16**b demonstrates a schematic of a solar TE cooler. In a solar TE cooler, the applied voltage of the cooling device comes from the solar panel. The device consists of two sides: the "cold side" and the "hot side." The cold side is placed inside the cooling compartment, such as a refrigerator or cooler, while the hot side is connected to metal fins that work as a heat sink to dissipate excess heat into the environment.[219] Solar TE coolers find applications in various products, including small refrigerators, wine coolers, portable beverage coolers, car coolers, personal air conditioning units, and more. They are particularly well-suited for electronic devices and small spaces where traditional compressor-based cooling systems might be impractical due to size constraints.[220]

TECs are excellent contenders for cooling electronics due to their scalable size, energy-efficient operation, and moving part-free configuration. The gradual decrease in chip sizes also requires a cooler of scalable size. Moreover, the local distribution of heat in microelectronics can be very small in size, which can be regarded as a hot spot. Recently, several studies have been underway to design effective hot spot coolers. Recently, Park et al. designed a filler-induced TEC device to eliminate the constraint of low thermal conductivity-induced low Fourier heat flow during hotspot cooling (**Figure 16**c).[221] They added more conductive filler material between the gap of the TE legs to better conduct the Fourier heat. A 10 K lower temperature could be obtained due to the introduction of fillers when compared to commercial TEC devices. Proper controlling of the TECs as hot spot coolers is significant in addition to material enhancement. Zhang et al. recently introduced the concept of using micro-TECs in an array configuration while controlling their individual operation to specifically cool down the area having hotspots.[222] Since hotspots can be generated at random locations in electronics, efficient control of TEC can also save energy. Better control and thermal management saved almost 34% of the energy for the TEC array when compared to the traditional TE cooler. The micro-TEC concept can also be extended for personal cooling requirements as well (**Figure 16**d, e).

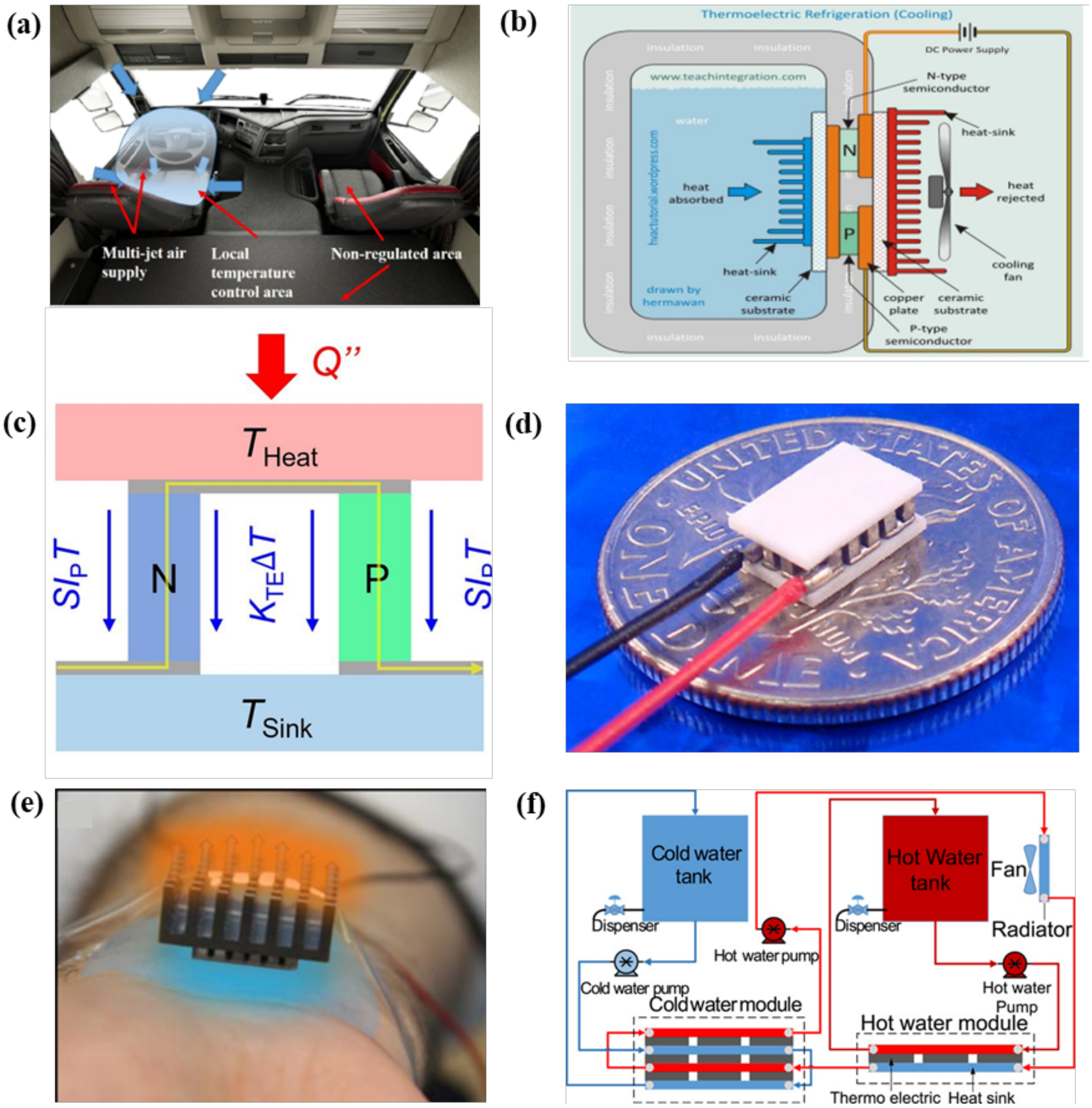

**Figure 16.** TE cooling applications. (a) Local climate control in a car. Reproduced with Permission.[211] Copyright 2022, Elsevier. (b) Solar TE refrigeration system. Reproduced with Permission.[219] Copyright 2023, The Authors. Published by Springer. (c) Concept of filler induced hotspot cooler. Reproduced with Permission.[221] Copyright 2023, The Authors. Published by Elsevier. (d) 3d printed micro-TEC, and (e) personal climate control. Reproduced with Permission.[222] Copyright 2023, Elsevier. (f) temperature-controlled water distribution system. Reproduced with Permission.[223] Copyright 2019, The Authors. Published by Wiley.

TE temperature management can be extended to water dispensers as well. The conventional water dispensing units with vapor compression cycle refrigeration system is too bulky for portability and less economic in taking up space. TEDs can be useful in this context with less energy consumption and moderate temperature control. In 2018 Hommalee et al. developed a temperature-controlled water dispensing system with TE plates.[223] A schematic of the water dispensing system is shown in **Figure 16**f. A typical system consists of two water tanks, two TE modules, three water pumps, and a radiator. The pumps continually feed water to the cold and hot sides of the TE module as required. A radiator is used on the hot side to cool the water temperature before feeding to the hot side TE module. Three cycles of water constitute the whole heating and cooling cycle of the system. A maximum hot-water temperature of 65 ºC and a minimum cold-water temperature of 13 ºC could be obtained in their study. In a follow-up study by the same group, Wiriyasart et al. modified the design of the fin configuration of the TE module for enhanced heat transfer.[224] They could achieve even lower cold-water temperatures of 8 ºC with modification. The modified setup also achieved the same level of power consumption when compared to the vapor compression-based water dispenser.

In conclusion, TE cooling technology has emerged as a transformative solution for efficient cooling across various industries. Its solid-state nature, precise temperature control, and potential for renewable energy integration make it a versatile and sustainable choice. By advancing materials, optimizing device design, and exploring innovative applications, TE cooling is poised to revolutionize how we approach cooling and energy efficiency challenges in the modern world.

## 5. Challenges and Future Perspectives

In recent years, significant progress has been made in the TE field to improve the efficiency of TE materials.[33,225] However, challenges remain to make thermoelectrics a commercially viable technology. This chapter will delve into the challenges and potential future directions of the TE field. This will include improved performance and scalability, cost reduction, thermal management, and integration with various energy systems.

### 5.1. Enhancing Thermoelectric Device Performances

Improving the performance of TEDs depends on two key aspects: improving the properties of the materials used and optimizing the device design and manufacturing process.[226,227]

### 5.1.1. Material Optimization

Enhancing the performance of TEDs greatly depends on optimizing TE material properties to achieve the increased figure of merit ZT. To acquire high ZT, in recent years, research related to the optimization of carrier concentration, the improvement of carrier mobility, and the reduction of its thermal conductivity have all been widely studied.[228–230] **Figure 17**a summarizes the ZT statistical data for several prominent TE materials from recent years.[231] The emergence of high-throughput material discovery technology has also promoted the latest advances in TE material optimization.[22,30] This high-throughput discovery process begins with the establishment of TE material libraries using various combinatorial processing techniques such as aerosol jet printing, co-sputtering, and pulsed laser deposition techniques. [22] The subsequent material libraries are rapidly evaluated for their TE properties, including electrical conductivity, thermal conductivity, and Seebeck coefficient. [31] The resulting large amount of data is quickly interpreted through data analysis and machine learning to screen out the optimized materials with desired properties. This high-throughput method has dramatically accelerated the discovery and development of TE materials with improved properties. [33]

Even though high-throughput TE material discovery offers great potential for identifying and optimizing new TE materials, several challenges persist. Future high-throughput TE material discovery methods will depend on continuous advances in computational techniques and machine learning algorithms. These developments are crucial for streamlining and speeding up the identification of high-performance materials.[232] Additionally, the development of cost-effective and rapid fabrication techniques is imperative. These methods will significantly expedite the creation of extensive material libraries.[233] Furthermore, a key research direction involves delving into the underlying mechanisms that drive TE performance, which is essential for laying a robust theoretical foundation for future TE material design.[234]

### 5.1.2. Device Design and Engineering

Achieving high-performing TEDs necessitates not just the selection of suitable materials but also the optimal TED design and engineering. These designs include the selection of fundamental TE conversion process directions (longitudinal and transverse TE effects), the choice of geometric shape and size (bulk, thin or thick film, cuboid or cylindrical), and the device's flexibility (flexible or rigid). Addressing factors such as the reliability and lifespan of TEDs is as crucial as the device design itself. Efficient use of a TED is unfeasible without adequately addressing contact-related challenges like mismatches of coefficients of thermal expansion (CTEs), chemical interactions, and mass diffusion.[235] Among the significant failure mechanisms in TEDs, the intense thermomechanical stress at the semiconductor-metal interface due to prolonged use and thermal cycling is arguably the most critical.[236] Furthermore, for some TEDs, such as thin-film devices, electrical and thermal contact resistances can often become dominant and significantly hinder device performance. The contact layer should minimize inherent parasitic electrical and thermal contact resistances[237,238] and ensure long-term chemical and electrical stability under harsh operating conditions. This is particularly important for TEDs that operate at high temperatures, as electrical contacts under high temperatures can easily become unstable and deteriorate over time.[239]

The emergent concept of functionally graded materials offers a new dimension to design TEDs by precisely engineering the local composition and properties of the TE materials, which can not only boost the overall efficiency of TEDs but also improve the interfacial properties between the TE materials and the metal contacts. As demonstrated by Bell et al., the maximum cooling temperature difference in a single-stage TED can be enhanced by utilizing distributed transport properties.[240,241] These functionally graded materials can also be applied to reduce contact resistances and thermal stresses due to the mismatches of thermal expansion coefficients. Taking advantage of their layer-by-layer manufacturing properties, recent advances in additive manufacturing provide an opportunity to manufacture these functionally graded materials.

Another strategy to improve the efficiency of TEDs is via segmented/cascaded module design that integrates multiple TE materials along the temperature gradient direction. The efficiency of segmented devices depends significantly on the ZT and compatibility factor of the individual materials in the segment and the contact resistances between the adjacent materials.[242] The segmented legs with low contact resistance can be formed either by consolidating all the materials at once, or joining the sintered individual segments.[243,244] Based on this concept, with nanostructured bismuth telluride and SKD materials, a solar TEG reached a top efficiency of 9.6% under an optical concentration of standard solar irradiance at 211 kW/m$^2$. [245] Similarly, an even higher efficiency of 11% has been showcased in the nanostructured segmented $Bi_2Te_3$/ PbTe module for a temperature difference of 590 K.[246]

## 5.2. Cost and Scalability

Cost and scalability are two critical factors in bringing TE technology out of the lab for widespread commercial use.[235,247] The former refers to the challenge of making these technologies economically viable for widespread use, while the latter is concerned with manufacturing techniques that can produce TEDs at a scale that meets increasing global demand.

### *5.2.1. Cost and Material Availability*

To date, a wide range of thermal management and waste heat recovery applications have been devised using TE materials. However, the high cost of TEDs (including material, manufacturing, and packaging costs) and the complex material supplies limit its commercialization and widespread application. High-performance TE materials are typically composed of rare or expensive elements and usually have a high cost.[248] In addition, the manufacturing process of TEDs is complex and requires precision, further increasing manufacturing costs. The problem with material supplies is twofold. First, the rare elements used in high-performance TE materials often lead to unstable supplies because of their scarcity. Second, the extraction and refining of these elements can pose significant environmental challenges, limiting their large-scale usage.[249,250]

Despite these challenges, reducing material consumption through innovative device structures and layouts and developing new sustainable materials offer viable ways to address existing challenges. First, the search for new cost-effective or Earth-abundant TE materials through high-throughput material discovery can potentially reduce material costs, harmful toxicity and negative environmental impacts, while providing high levels of energy conversion efficiency and considerable power output.[251,252] Furthermore, refining the structure of the device[253] (**Figure 17**c) and the configuration of the TE modules[254] could lead to decreased material usage.

### *5.2.2. Manufacturing Techniques*

In order to achieve the ideal material properties and device efficiency, TE materials and devices often need to be carefully synthesized and processed. However, traditional manufacturing methods often involve complex and expensive manufacturing processes that affect their commercial viability and large-scale production and deployment. For example, the production of metal chalcogenides, which are highly efficient TE materials, requires intricate steps such as ball milling, quenching, annealing, and spark plasma sintering. [255] Moreover, the typical non-planar nature of heat sources limits the usefulness of TEGs made from standard bulk materials (like BiTe-, SnSe-, or PbTe-based alloys). These materials are created through commonly employed techniques such as hot pressing,[256] spark plasma sintering,[257] and zone melting,[258] resulting in rigid bulk materials and devices, which restrict their applicability. Therefore, there is an urgent need to develop new manufacturing techniques to produce TED with non-planar or irregular complex shapes using scalable and low-cost processes.

Additive manufacturing, especially ink-based printing processes, is emerging as a potential solution for scalable manufacturing of TE materials with non-planar or irregular complex shapes. Several printing techniques, such as screen printing, inkjet printing, aerosol jet printing, and extrusion printing (**Figure 17**d), have been utilized in the production of TE materials and devices of almost any shape, [28,61] as elaborated in **Chapter 3**. However, the additive manufacturing of TE materials is still in its early stages, and the performance of printed materials needs to be continuously improved by carefully controlling the composition of TE material-based inks, optimizing the printing parameters and paths, and improving the device post-printing processing and sintering.[61]

High-throughput production of TEDs also necessitates rapid sintering in ink-based printing techniques. Traditional methods for fabricating bulk TE materials typically involve high-temperature processes to sinter TE powders into dense structures, such as spark plasma sintering and hot pressing, either individually or in combination. However, the sintering processes used in these processes are often complex and require specialized and expensive equipment. Some of these processes are also incompatible with layer-by-layer deposition techniques, as they may damage pre-existing structures. To facilitate rapid manufacturing of both bulk and thin-film TEDs, several alternative sintering techniques are currently under investigation, including pulsed light sintering[59] and flash spark plasma sintering.[86] However, comprehensive studies are still needed to understand how these processes alter the TE materials' microstructure and transport properties. Ongoing research in the sintering techniques for TE materials could significantly boost the efficiency of TEDs, paving the way for the evolution of more effective and eco-friendly energy conversion solutions.

### 5.3. Hybrid Energy Harvesting and Thermal Management

Two main approaches have come to the forefront to improve the overall efficiency of the TE system: integrating hybrid energy harvesting systems and introducing advanced thermal management. Hybrid energy harvesting systems combine TEDs with other energy systems, such as solar or wind.[259,260] Advanced thermal management designed to improve TED efficiency by fine-tuning heat transfer paths and temperature levels.[261] With the synergies of various approaches, the overall efficiency of TEDs is gradually improving, making it more likely to become an attractive solution for multiple applications.

#### *5.3.1. Hybrid Energy Systems*

Combining TE technology with other power generation solutions to create hybrid energy systems can greatly reduce costs and improve overall efficiency, compared to only using TEDs for power generation.[262] TEGs have been combined with a variety of renewable energy systems, such as solar, fuel cells, and biomass, to improve energy efficiency.[263] In addition to this, the integration of TE and triboelectric nanogenerators has also been reported, and this hybrid energy harvesting system is lightweight, low-cost, and efficient, making it ideal for powering devices in the Internet of Things (IoT) (**Figure 17**e).[264]

However, some obstacles still limit the successful integration of TEGs into hybrid systems. First, the control and management mechanisms of hybrid energy systems (HES) are complex. In order to achieve the perfect integration of TEGs with other power generation systems and improve efficiency, it is necessary to develop better control and management mechanisms.[265] In addition, it is essential to ensure the thermal stability of the TE material and its connection with other components in the hybrid system.[235] Moreover, the unreliability and weather dependence of many renewable energy sources is a significant hurdle in HES. When these energy sources are integrated with TE technology, this dependence on weather or supply instability can become a problem, as it can affect the overall stability and efficiency of the energy system.[265]

While obstacles remain in incorporating TEGs into hybrid energy systems, the future for these systems looks bright. Leveraging artificial intelligence (AI) or machine learning can enhance the management, optimization, and control of these hybrids, thus boosting their effectiveness and dependability.[265] Furthermore, developing hybrid TEs, which integrate organic and inorganic materials at different scales, could lead to performance enhancements that enable successful integration with HES.[266]

*5.3.2. Thermal Management*

To amplify the performance of TEDs, thermal management is of the utmost importance.[267] Controlling and optimizing the thermal flow through these materials can greatly affect their efficiency.[268] Several advanced strategies have been developed for improved thermal management. These include meticulously designing the heat exchanger's shape for peak effectiveness,[269] innovatively altering its internal architecture to facilitate better heat transfer,[261] and introducing superior cooling methods[270] (**Figure 17**f).

In order to improve the efficiency of TEDs through proper thermal management, a comprehensive grasp of heat transfer and fluid dynamics (mainly the transfer of thermal energy within and between various media) is essential. A common technique utilized to analyze and comprehend these mechanisms is the Finite Element Method (FEM), which can enable optimization of complex designs through simulations without experiments.[194,271] In addition to these efforts in principles, the integration of Phase Change Materials (PCMs) has gained traction in enhancing TED system performance. PCMs are a class of substances that can trap and release heat when undergoing phase transitions, such as melting and freezing. By combining the latent heat storage capacity of PCMs for thermal management, the heat transfer rate of TEDs can be further improved.[272,273]

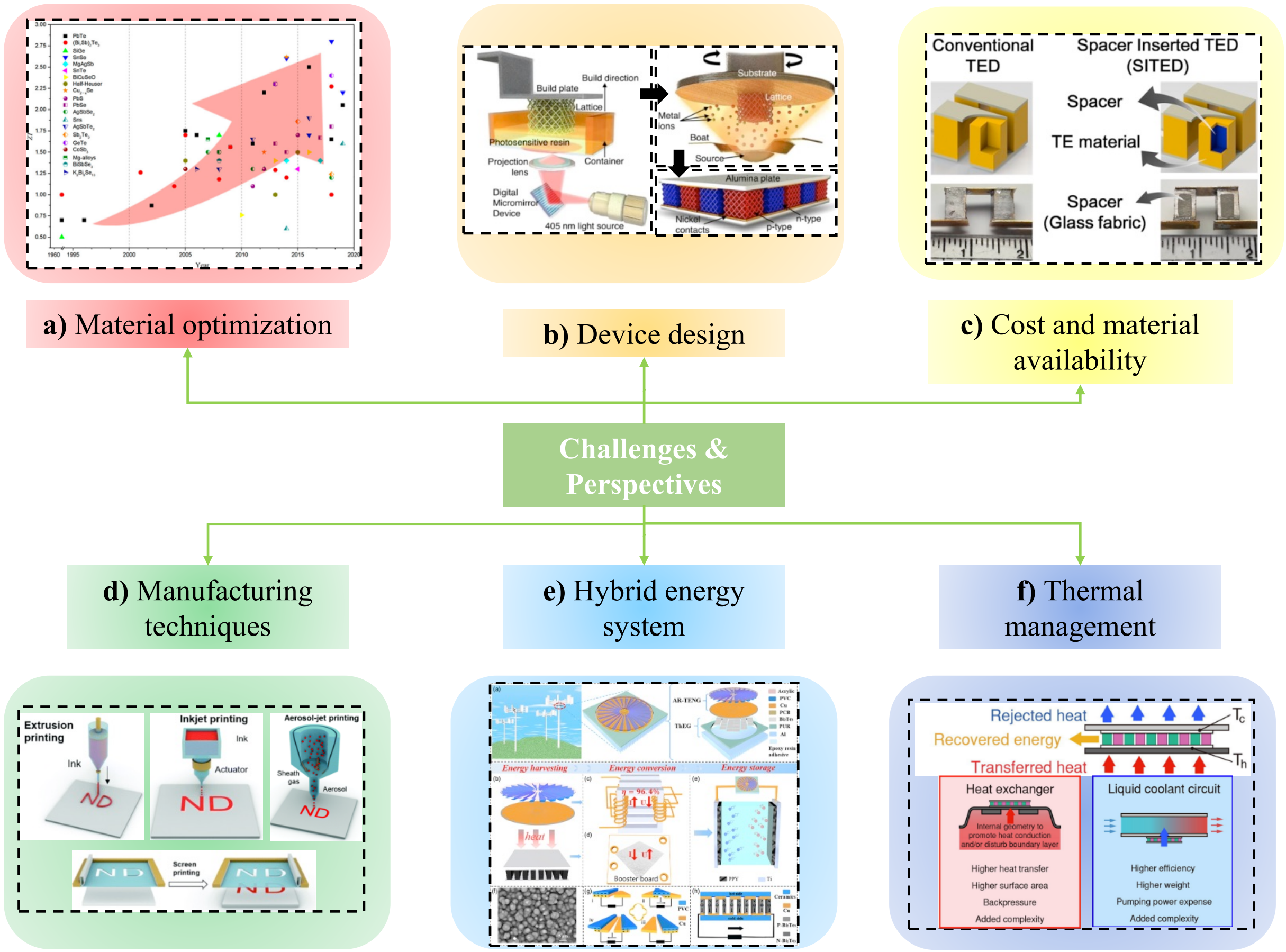

**Figure 17.** Challenges & Perspectives of TED industry. (a) Material optimization, Reproduced with Permission.[231] Copyright 2022, Frontiers. (b) Device design, Reproduced with Permission.[227] Copyright 2023, Springer Nature. (c) Cost and material availability, Reproduced with Permission.[253] Copyright 2017, Elsevier. (d) Manufacturing techniques, Reproduced with Permission.[28] Copyright 2022, The Authors. Royal Society of Chemistry. (e) Hyrbid energy system, Reproduced with Permission.[264] Copyright 2022, Published by Wiley. (f) Thermal management. Reproduced with Permission.[268] Copyright 2021, Elsevier.

### 5.4. Conclusions

This review provides a comprehensive roadmap for the development of TE technology, from material discovery and device fabrication to practical applications. High-throughput materials discovery methods that combine material synthesis/characterization/computational techniques (machine learning/advanced simulation, etc.) effectively explore a wide array of materials, leading to rapid identification of innovative TE compounds as well as increased material discovery efficiency. In addition, the scalable manufacturing process for cost-effective, high-performance TEDs is discussed in terms of ink-based printing processes, rapid sintering techniques, and device contact processes. Furthermore, the applications of TEDs in waste heat recovery and thermal management systems are also introduced. Despite significant progress in this area, continuous research efforts are needed to achieve widespread application and commercialization of TEDs in optimizing material properties, improving device efficiency, reducing equipment costs, and hybrid system integration.

**Acknowledgements**

K. S. and A. N. M. T. contributed equally to this work. The authors would like to acknowledge funding support from the U.S. Department of Energy under awards DE-EE0009103 and DE-NE0009138, and the National Science Foundation under award CMMI-1747685.

**Conflict of Interest**

The authors declare no conflict of interest.